\documentclass[preprint,12pt]{elsarticle}

\usepackage{amssymb}

\journal{Nuclear Physics B}

\begin{document}

\begin{frontmatter}

\title{Search for lepton flavor violation in supersymmetric models via meson decays}

\author[label1,label2]{Ke-Sheng Sun \corref{cor1}}
\ead{sunkesheng@126.com}
\cortext[cor1]{Corresponding author}
\author[label1,label2]{Tai-Fu Feng  }
\author[label1,label2]{Tie-Jun Gao  }
\author[label1]{Shu-Min Zhao }

\address[label1]{Department of Physics, Hebei University, Baoding 071002,China}
\address[label2]{Department of Physics, Dalian University of Technology,Dalian 116024, China }

\begin{abstract}
Considering the constraints from the experimental data on $\mu\rightarrow e\gamma$, $\mu\rightarrow3e$,
$\mu-e$ conversion etc., we analyze the Lepton Flavor Violating decays $\phi(J/\Psi,\Upsilon(1S))
\rightarrow e^+\mu^-(\mu^+\tau^-)$ in the scenarios of the minimal supersymmetric extensions of
Standard Model with seesaw Mechanism. Numerically, there is parameter space
that the LFV processes of $J/\Psi(\Upsilon)\rightarrow\mu^+\tau^-$ can reach the upper
experimental bounds, meanwhile the theoretical predictions on $\mu\rightarrow e\gamma$, $\mu\rightarrow3e$,
$\mu-e$ conversion satisfy the present experimental bounds. For searching of new physics, Lepton Flavor Violating processes $J/\Psi(\Upsilon)\rightarrow\mu^+\tau^-$ may be more promising and effective channels.
\end{abstract}

\begin{keyword}
Lepton flavor violating \sep supersymmetry \sep see-saw.


\MSC[2010] 81T60 \sep 81V15
\end{keyword}

\end{frontmatter}


\section{Introduction}
\label{intro}
As an evidence to discover new physics beyond the Standard Model (SM), searching for Lepton Flavor Violating (LFV) processes
in charged lepton sector have attracted a great deal of attention. The theoretical predictions on those lepton flavor
violating processes are suppressed by small masses of neutrinos in SM, and exceed the detecting extent
of experiment in near future. Nevertheless, the corrections to the branching ratios of LFV decays
$\phi\rightarrow e^+\mu^-$, $J/\Psi\rightarrow\mu^+\tau^-$ and $\Upsilon\rightarrow\mu^+\tau^-$ are
enhanced by the new sources of LFV in various extensions of the SM, such as grand unified models \cite{GUT},
supersymmetric models with and without R-parity \cite{SUSY}, left-right symmetry models \cite{LR} etc.
Although nonzero neutrino masses supported by the neutrino oscillation experiments \cite{oscillation}
imply the non-conservation of lepton flavor, it is very important to directly search the LFV processes
of charged lepton sector in colliders running now.

Using a sample of $5.8\times10^7\;J/\Psi$ events collected with the BESII detector, Ref.\cite{BESII} obtains the upper limits on BR$(J/\Psi\rightarrow\mu\tau)<2.0\times10^{-6}$ and BR$(\Upsilon\rightarrow\mu\tau)<8.3\times10^{-6}$ at the 90\% confidence level (C.L.). Adopting the data collected with the CLEO III detector, the authors of Ref.\cite{CLEO} estimate the upper limits on BR $(\Upsilon(1S)\rightarrow\mu\tau)<6.0\times10^{-6}$, BR $(\Upsilon(2S)\rightarrow\mu\tau)<1.4\times10^{-5}$ and BR $(\Upsilon(3S)\rightarrow\mu\tau)<2.0\times10^{-5}$ respectively at the  95\% C.L. Additionally, the study of LFV
processes involving light unflavored meson is an effective way maybe to search for new physics beyond the SM, and the SND Collaboration at the BINP (Novosibirk) presents an upper limit on the $\phi\rightarrow e^+\mu^-$ branching fraction of BR $(\phi\rightarrow e^+\mu^-)\le 2\times 10^{-6}$ \cite{SND}.

In literature, several stringent limits on LFV decays of both light and heavy unflavored mesons are derived already. Assuming that a vector boson $M_i$ ($M_i$ could be either a fundamental state, like the $Z_0$, or a quark-antiquark bound state like the $\phi,\;J/\Psi,\;\Upsilon$) couples to $\mu^{\mp}e^{\pm}$ and $e^{\mp}e^{\pm}$ as:
\begin{eqnarray}
{\cal L}_{eff}=g_{_{M_i\mu e}}\bar{\mu}\gamma_\mu eM^\mu_i
+g_{_{M_iee}}\bar{e}\gamma_\mu eM^\mu_i+h.c.\;,
\end{eqnarray}
where $g_{_{M_iee}}$ and $g_{_{M_i\mu e}}$ denote the corresponding couplings of a meson to lepton flavor conservation and violation currents, and by unitarity its exchange contributes to $\mu\rightarrow3e$, the authors of Ref.\cite{zhang} deduce upper bounds
on the LFV decay of mesons from the LFV process $\mu\rightarrow3e$.
Under a similar assumption that a vector meson $M_i$ couples to $\mu^{\mp}e^{\pm}$ and $N N$ as:
\begin{eqnarray}
{\cal L}_{eff}=(\xi^{M}_V\bar{e}\gamma_\mu \mu +\xi^{M}_A\bar{e}\gamma_\mu \gamma_5\mu)M^\mu_i
+g_{_{M NN}}\bar{N}\gamma_\mu N M^\mu_i+h.c.\;,
\end{eqnarray}
where N is a nucleon, $\xi^{M}_{V,A}$ are effective vector and axial couplings of a meson to the LFV lepton currents, authors of Ref.\cite{Gutsche} studies the LFV decays of vector mesons by taking account of the experimental constraint on $\mu-e$ conversion. It shows the constraint from $\mu-e$ conversion on LFV decays of vector mesons is more stronger. Likewise, authors of Ref.\cite{zhang} also deduce upper bounds on other LFV decay of mesons from the LFV processes $\tau\rightarrow3e$ and $\tau\rightarrow3\mu$. Making the assumption that fermion mixing and mass hierarchy originate from mass matrix rotation, authors of Ref.\cite{Bordes} also get some upper limits on the LFV decays of heavy unflavored mesons and Z boson. Searching for new physics beyond the SM is also a goal of LHC. In LHC, vector mesons can be produced by photo fusion \cite{Armestol}.

In SM,the LFV decays mainly originate from the charged current with the mixing among three lepton generations.
The fields of the flavor neutrinos in charged current weak interaction Lagrangian are combinations of three massive neutrinos:
\begin{eqnarray}
{\cal L}& = &-\frac{g_2}{\sqrt{2}}\sum_{l=e,\mu,\tau}\overline{l_{L}}(x)\gamma_\mu \nu_{lL}(x)W^\mu(x)+h.c.,
\nonumber\\
\nu_{lL}(x)& = &\sum_{i=1}^{3}\Big(U_{MNS}\Big)_{li}\nu_{iL}(x),
\end{eqnarray}
where $g_2$ denotes the coupling constant of gauge group SU(2), $\nu_{lL}$ are fields of the flavor neutrinos, $\nu_{iL}$ are fields of massive neutrinos,
and $U_{MNS}$ corresponds to the MNS neutrino mixing matrix \cite{Pontecorvo,Maki}.In the standard parametrization \cite{PDG}, the leptonic mixing matrix is given by:
\begin{eqnarray}
U_{MNS}
&=&\left(\begin{array}{ccc}
c_{1}c_{3}&c_{3}s_{1}&s_{3}e^{-i\delta}\\
-c_{1}s_{3}s_{2}e^{i\delta}-c_{2}s_{1}&c_{1}c_{2}-s_{1}s_{2}s_{3}e^{i\delta}&c_{3}s_{2}\\
s_{1}s_{2}-c_{1}s_{3}c_{2}e^{i\delta}&c_{1}s_{2}-s_{1}c_{2}s_{1}e^{i\delta}&c_{3}c_{2}
\end{array}\right)\nonumber\\
&&\times   diag\Big(e^{i\Phi_{1}/2},1,e^{i\Phi_{2}/2}\Big),
\label{MNS}
\end{eqnarray}
where $s(c)_{1}$ = $\sin(\cos)\theta_{12}$, $s(c)_{2}$ = $\sin(\cos)\theta_{23}$, $s(c)_{3}$ = $\sin(\cos)\theta_{13}$.
The phase $\delta$ is the Dirac CP phase, and $\Phi_{i}$ are the Majorana phases. A global fit of the neutrino
oscillation data points out: $\theta_{12}\sim 34^{\circ}$ and $\theta_{23}\sim 45^{\circ}$.
Recently, the observing $\bar{\nu}_e$ disappearance in reactor experiments Daya Bay \cite{Dayabay} and RENO \cite{Reno}
have definitely established that $\theta_{13}>0$ at $\sim5\sigma$ level. The Daya Bay and RENO have measured
$\sin \theta_{13}\simeq0.024$ and $\sin \theta_{13}\simeq0.029$ , respectively.
However, the theoretical predictions on branching ratios of any LFV decays are suppressed strongly by the
tiny neutrino masses in SM and fall out the reach of experiment in near future. In this work, we analyze the LFV decays :
$\phi\rightarrow e^+\mu^-$, $J/\Psi\rightarrow\mu^+\tau^-$ and $\Upsilon(1S)\rightarrow\mu^+\tau^-$ in the
framework of accomodating supersymmetry with type I seesaw mechanism simultaneously.
With the accumulation of events on BEPC \cite{BESII} and SuperKEKB \cite{SuperKEKB},
the updated experimental data on those LFV decays maybe constrain the concerned models more stringent.
To shorten the length of text, we just present the upper bounds on those branching ratios of
$\rho (\omega, J/\Psi, \Upsilon)\rightarrow e^+ \mu^-$ under our assumptions on parameter space.

The paper is organized as follows. In Section.\ref{sec:2}, we firstly provide a simple overview for the origin
of lepton flavor changing and corresponding interaction lagrangian in the framework of MSSM with type I seesaw mechanism.
Then, as an example,we derive the analytic results of amplitude for one diagram in detail. The numerical results are presented
in Section.\ref{sec:3}, and the conclusion is drawn in Section.\ref{sec:4}.
All the simplified amplitudes corresponding to the Feynman diagrams in Fig.\ref{fig1} and Fig.\ref{fig2} are given in \ref{app}.

\section{Formalism}
\label{sec:2}
In the minimal supersymmetric extension of SM with R-parity conservation, the general form of the superpotential involving the lepton and
Higgs superfields is given by \cite{Rosiek}:
\begin{eqnarray}
{\cal W}_{MSSM}&=&\epsilon^{ij}\Big(\mu\hat{H}^1_i\hat{H}^2_j+Y_l^{IJ}\hat{H}^1_i\hat{L}^I_j
\hat{R}^J\Big),
\label{MSSM-superpotential}
\end{eqnarray}
where $\mu$ is the mu-parameter, the $3\times3$ matrix $Y_l$ is the charged lepton Yukawa couplings.
For convenience, we assume $Y_{l}^{IJ}=Y_{l}^I\delta^{IJ}$ (I,J=1,2,3) in this work.
Then, the relevant soft supersymmetry breaking terms involving the slepton sector and sneutrino sector are:
\begin{eqnarray}
V^{MSSM}_{soft}&=&\Big(m_{_{L}}^2\Big)^{IJ}\tilde{L}_i^{I*}\tilde{L}_i^J+\Big(m_{_{R}}^2\Big)^{IJ}\tilde{R}^{I*}\tilde{R}^J
-A_l^{IJ}\epsilon^{ij}H^1_i\tilde{L}^I_j\tilde{R}^J
\nonumber\\
&&-{{A}'_l}^{IJ}H^{2*}_i\tilde{L}^I_i\tilde{R}^J-h.c.\;,
\label{MSSM-soft-term}
\end{eqnarray}
where $m_{L}^2$ is left $3\times3$ soft slepton mass matrix, $m_{R}^2$ is right $3\times3$ soft slepton
mass matrix, the $3\times3$ matrix $A_{l}$ is the trilinear scalar couplings, the $3\times3$ matrix
${A}'_{l}$ is the non-standard trilinear scalar couplings, respectively.
The LFV interactions mainly originate from the potential misalignment between the leptons and sleptons
mass matrices in the MSSM. In other words, the sources of LFV are the off-diagonal entries of the $3\times3$
soft supersymmetry breaking matrices $m_{L}^{2}$, $m_{R}^{2}$, $A_{l}$ and ${A}'_l$ in
$6\times6$ slepton mass matrix $M^2_{L}$, which are listed below:
\begin{eqnarray}
&&\Big(M_{L}^{2}\Big)_{LL}^{IJ} =\frac{e^{2}(\upsilon _{1}^{2}-
\upsilon _{2}^{2})( 1-c_{\rm w}^{2})}{8s_{\rm w}^{2}c_{\rm w}^{2}}\delta^{IJ}
+\frac{\upsilon _{1}^{2}(Y_{l}^I)^{2}}{2}\delta^{IJ}
+(m_{L}^{2})^{JI}\;,\\
&&\Big(M_{L}^{2}\Big)_{RR}^{IJ} = -\frac{e^{2}(\upsilon _{1}^{2}
-\upsilon _{2}^{2})}{4c_{\rm w}^{2}}\delta^{IJ}+\frac{\upsilon _{1}^{2}(Y_{l}^I)^{2}}{2}\delta^{IJ}
+(m_{R}^{2})^{IJ}\;,\\
&&\Big(M_{L}^{2}\Big)_{LR}^{IJ}=\frac{1}{\sqrt{2}}
(\upsilon _{2}(\mu ^{\ast }Y_{l}^I\delta^{IJ}-{A}_{l}^{\prime IJ})+\upsilon _{1}A_{l}^{IJ})\;,
\label{eq2}
\end{eqnarray}
where $s_{\rm w}$=$\sin\theta _{\rm w}$,$c_{\rm w}$=$\cos\theta _{\rm w}$ with $\theta _{\rm w}$ denoting the Weinberg angle,
and $\upsilon_{1,2}$ are the non zero vacuum expectation values (VEVs) of two Higgs doublets.

In the minimal supersymmetric extension of the seesaw extended SM \cite{Rosiek2,SeeSaw1,SeeSaw2,SeeSaw3,SeeSaw4,SeeSaw5},
there are three generation right handed neutrino superfields $\hat{N}^I\;(I=1,\;2,\;3)$ with zero hypercharge.
The most general form of the superpotential involving the lepton and Higgs superfields
in the R-parity conserving scenario is given by:
\begin{eqnarray}
{\cal W}&=&{\cal W}_{MSSM}+\epsilon^{ij}Y_\nu^{IJ}\hat{H}^2_i\hat{L}^I_j\hat{N}^J+{1\over2}M^{IJ}\hat{N}^I\hat{N}^J,
\label{SeeSaw-superpotential}
\end{eqnarray}
where $Y_\nu$ is the $3\times3$ neutral lepton Yukawa coupling, $M$ is the $3\times3$ Majarana mass matrix. Here, we adopt the parameterization in \cite{Rosiek2}
to reproduce the PMNS mixing matrix:
\begin{eqnarray}
(Y_{\nu})^{ij}=i\sum_{k=1}^{3}\sqrt{2}(m_{\nu^k_L}M_{\nu^j_R})^{1/2}R_{jk}(U^{\ast}_{MNS})_{ik}/\upsilon_{2}\;,
\end{eqnarray}
where $U_{MNS}$ is the MNS mixing matrix in Eq.(\ref{MNS}), $m_{\nu^i_L} (i= e,\mu,\tau)$ are the masses
of left handed neutrinos, and $M_{\nu^i_R}(i = e,\mu,\tau)$ are right handed neutrino masses.
Furthermore, R is an arbitrary orthogonal matrix \cite{Casas} determined by
three angles $\alpha_{1}$, $\alpha_{2}$, $\alpha_{3}$:
\begin{equation}
R=\left(\begin{array}{ccc}
c_{2} c_{3} &-c_{1} s_{3}-s_{1} s_{2} c_{3}  &s_{1} s_{3}-c_{1} s_{2} c_{3} \\
c_{2} s_{3} &c_{1} c_{3}-s_{1} s_{2} s_{3}  &-s_{1} c_{3}-c_{1} s_{2} s_{3} \\
 s_{2} &s_{1} c_{2}  &c_{1} c_{2}
\end{array}\right),\nonumber
\end{equation}
in which $c_{i}=\cos\alpha_{i}$ and $s_{i}=\sin\alpha_{i}$, $i=1, 2, 3$. In the scenarios of MSSM with Seesaw mechanism,
the corrections from right handed Majorana neutrinos to the branching ratios of vector meson LFV decays can
be ignored since they are suppressed by the huge masses of right handed neutrinos.
In addition, the mass term for the light sneutrinos is given by:
\begin{eqnarray}
&&-{\cal L}_{\tilde\nu}^{mass}={1\over2}\left(\begin{array}{cc}\tilde{\nu}_L^I,\;&\tilde{\nu}_L^{I*}\end{array}\right)
{\cal M}^2_{\tilde \nu}\left(\begin{array}{c}\tilde{\nu}_L^J \\\tilde{\nu}_L^{J*}\end{array}\right),
\end{eqnarray}
with $I,\;J=1,\;2,\;3$ are the indices of generation, and the $6\times6$ mass matrix is
\begin{eqnarray}
&&{\cal M}^2_{\tilde \nu}=
\left(\begin{array}{cc}(M^{2}_{LC})^{IJ}&(M^{2*}_{LV})^{IJ}\\(M^{2}_{LV})^{IJ}&(M^{2*}_{LC})^{IJ}
\end{array}\right).
\end{eqnarray}
Here $M^{2}_{LC}$ and $M^{2}_{LV}$ are $3\times3$ matrices. If $M^{2}_{LV}=0$, the six light sneutrinos
are comprised of three sneutrino-antisneutrino pairs. If $M^{2}_{LV}\ne 0$, the lepton number is
violated and the sneutrinos and antisneutrinos can mix and yield six non-degenerate sneutrinos.
The elements of $M^{2}_{LC}$ and $M^{2}_{LV}$ are given by, in a simple form at GUT scale:
\begin{eqnarray}
&&(M^2_{LC})^{IJ}=(m^2_{L})^{IJ}+\frac{1}{2}M^2_{Z} \cos2\beta\delta^{IJ}, \\
&&(M^2_{LV})^{IJ}=-\frac{(\upsilon_2 )^2\mu^{*} \cot\beta}{2}(Y_{\nu}M^{-1}Y_{\nu}^{T})^{IJ} ,
\end{eqnarray}
where $M$ is the right handed neutrino mass matrix in Eq.(\ref{SeeSaw-superpotential})
and $\tan\beta=\upsilon_2/\upsilon_1$. $M^{2}_{LC}$ reproduces the well known $3\times3$ light sneutrino matrix in MSSM.
In the CP-base
\begin{eqnarray}
&&\tilde{\nu}_L^{(+)I}={1\over\sqrt{2}}\Big(\tilde{\nu}_L^I+\tilde{\nu}_L^{I*}\Big), \\
&&\tilde{\nu}_L^{(-)I}=-{i\over\sqrt{2}}\Big(\tilde{\nu}_L^I-\tilde{\nu}_L^{I*}\Big) ,
\end{eqnarray}
the mass term for the light sneutrinos is rewritten as:
\begin{eqnarray}
&&-{\cal L}_{\tilde\nu}^{mass}={1\over2}\left(\begin{array}{cc}\tilde{\nu}_L^{(+)I},\;&\tilde{\nu}_L^{(-)I}\end{array}\right)
\bar{\cal M}^2_{\tilde \nu}\left(\begin{array}{c}\tilde{\nu}_L^{(+)J} \\\tilde{\nu}_L^{(-)J}\end{array}\right)\;.
\end{eqnarray}
Here, the $6\times6$ mass-squared matrix $\bar{\cal M}^2_{\tilde \nu}$ is
\begin{eqnarray}
\bar{{\cal M}}^2_{\tilde \nu}=
{\small\left(\begin{array}{cc}
\Re\Big((M^{2}_{LC})^{IJ}+(M^{2}_{LV})^{IJ}\Big)&-\Im\Big((M^{2}_{LC})^{IJ}+(M^{2}_{LV})^{IJ}\Big)\\
\Im\Big((M^{2}_{LC})^{IJ}-(M^{2}_{LV})^{IJ}\Big)&\Re\Big((M^{2}_{LC})^{IJ}-(M^{2}_{LV})^{IJ}\Big)
\end{array}\right)}\;.
\end{eqnarray}
The effective squared-mass matrix can be diagonalized by $6\times6$ orthogonal matrix, $Z_{\tilde \nu}$ via:
\begin{eqnarray}
Z^T_{\tilde \nu}\bar{{\cal M}}^2_{\tilde \nu}Z_{\tilde \nu}=(m^2_{S_1}, m^2_{S_2}, ...,m^2_{S_6}),
\end{eqnarray}
where $S_i(i=1, ..., 6)$ correspond to the physical sneutrino mass eigenstates. The sneutrino interaction eigenstates,
$\tilde {\nu}^I$, can be expressed in terms of the physical sneutrino mass eigenstates $S_k$ by:
\begin{eqnarray}
\tilde {\nu}_L^I=\frac{1}{\sqrt {2}}\sum_{k=1}^{6}(Z^{Ik}_{\tilde {\nu}}+iZ^{(I+3)k}_{\tilde {\nu}})S_k,
\end{eqnarray}

\begin{figure}
\centering
\includegraphics[width=3.0in]{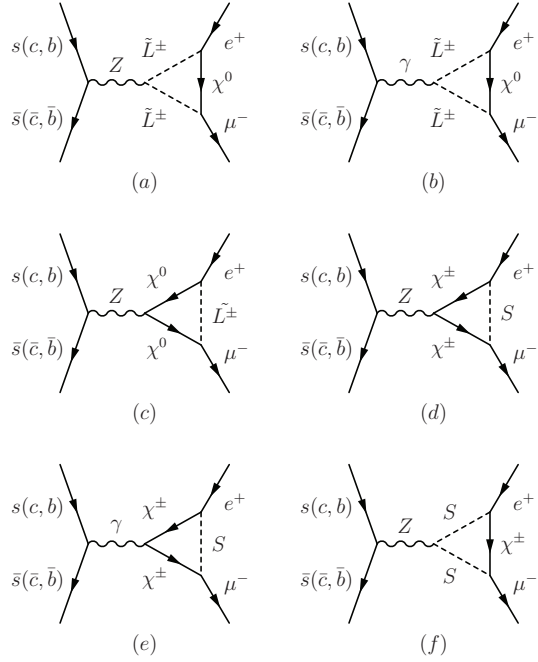}
\caption[]{The penguin diagrams of the LFV processes $\phi(J/\Psi,\Upsilon)\rightarrow e^+\mu^-(\mu^+\tau^-)$
in MSSM with seesaw mechanism.}
\label{fig1}
\end{figure}

\begin{figure}
\centering
\includegraphics[width=3.0in]{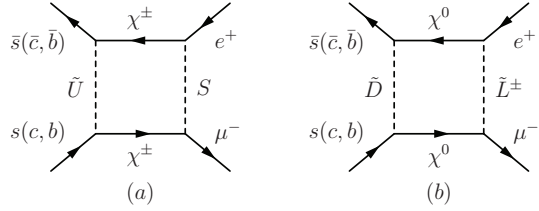}
\caption[]{The box diagrams of the LFV processes $\phi(J/\Psi,\Upsilon)
\rightarrow e^+\mu^-(\mu^+\tau^-)$ in the MSSM with seesaw mechanism.
In concrete calculation, the corrections from cross diagrams should be included also.}
\label{fig2}
\end{figure}

Correspondingly, the relevant Lagrangian is given as:
\begin{eqnarray}
{\cal L}&=&\bar{\chi }_{j}^{0}\Big[\Big(\frac{e}{\sqrt{2}s_{\rm w}c_{\rm w}}Z_{L}^{Ii}
( Z_{N}^{1j}s_{\rm w}+Z_{N}^{2j}c_{\rm w})+Y_{l}^{I}Z_{L}^{(J+3)i} Z_{N}^{3j}\Big)P_{L}
\nonumber\\&&+ \Big(
\frac{-e\sqrt{2}}{c_{\rm w}}Z_{L}^{\left (I+3  \right )i}Z_{N}^{1j\ast }
+Y_{l}^{I}Z_{L}^{Ii} Z_{N}^{3j\ast } \Big)P_{R}\Big ]e^{I}{\tilde L}_{i}^{+}
\nonumber\\&&-\bar{\chi }_{i}^{C}\left (\frac{e}{s_{\rm w}}Z_{+}^{1i}P_{L}+Y_{l}^{I}
Z_{-}^{2i\ast }P_{R} \right )
\Big(Z_{\tilde{\nu} }^{Ij}-iZ_{\tilde{\nu} }^{(I+3)j}\Big)e^{I}S_{j }+h.c.,
\end{eqnarray}
where $Z_{\pm}$ are the mixing matrices of chargino sector, $Z_{N}$, $Z_{L}$ and $Z_{\tilde{\nu}}$ are the mixing matrices of neutralino sector, slepton sector and sneutrino sector, respectively. $e^I$ denote the SM charged leptons. ${\tilde L}_{i}^{+}$ and $S_{j}$ denote the sleptons and sneutrinos. $\chi_{i}^{C}$ and $\chi_{j}^{0}$ stand for the charginos and neutralinos. $P_{L/R} = \frac{1}{2}(1\mp\gamma_{5})$. The relevant Feynman diagrams contributing to the LFV decays are presented in Fig.\ref{fig1} and Fig.\ref{fig2}.

In the quark picture, mesons are composed of a quark and an anti-quark. As we analyze
those LFV processes mentioned above, we do not want to calculate the complicated loop
integrations at quark-gluon level since the lack of a completely reliable way to
calculate the non-perturbative QCD effects. We adopt a phenomenological model
where the amplitude of hard process involving a s-wave meson
can be described by the matrix elements of gauge invariant nonlocal operators,
which are sandwiched between the vacuum and the meson states. For our case,
the matrix is given by \cite{OZI}:
\begin{eqnarray}
\langle 0| \bar{q}(y) \Gamma[y,x] q(x)|   \phi \rangle\;,
\end{eqnarray}
here $\Gamma=\gamma_{\mu}$ or $\gamma_{\mu}\gamma_{5}$ is a generic Dirac matrix structure,
$x$ and $y$ are the coordinates of quark and anti-quark. The distribution amplitude of vector meson $\phi$
in leading-order is defined through the correlation function :
\begin{eqnarray}
\langle 0| \bar{s}_{1\alpha }^{i}(y)s_{2\beta}^{j}(x)|\phi(p)\rangle
&=&\frac{\delta_{ij}}{4N_{c}}\int_{0}^{1}due^{-i(upy+\bar{u}px)}
\Big[ f_{\phi }m_{\phi }/\!\!\!\varepsilon _{\phi }\phi _{\parallel}(u)
\nonumber\\&&+\frac{i}{2}\sigma ^{{\mu }'{\nu }'}f_{\phi }^{T}
\Big( \varepsilon _{\phi {\mu }'}{p}_{{\nu }'}-\varepsilon _{\phi {\nu }'}{p}_{{\mu }'}\Big
)\phi _{\perp } ( u )\Big]_{\beta \alpha}\;,
\label{hadron}
\end{eqnarray}
where the momentum of $\phi$ is on-shell, i.e. $p^{2}=m_{\phi}^{2}$,
$\varepsilon_{\phi }$ is the polarization vector, $f_{\phi }$ and $f_{\phi }^{T}$ are
the decay constants of $\phi$ meson, $\phi _{\parallel }$ and $\phi _{\perp }$ are the
leading-twist distribution functions corresponding to the longitudinally
and transversely polarized meson, respectively.
For the cases of $J/\Psi$ and $\Upsilon$, there are similar distribution amplitudes.
The integration variable u corresponds to the momentum fraction carried by the quark, $\bar{u}=1-u$
stands for the momentum fraction carried by the anti-quark, $\alpha$ and $\beta$ are the indices of matrix elements,
and Nc is the number of colors, separatively. Since the leading-twist light-cone
distribution amplitudes of meson are close to their asymptotic form \cite{Beneke},
so we set $\phi _{\parallel }=\phi _{\perp }=\phi(u)=6u(1-u)$.
\begin{figure}
\begin{center}
\includegraphics[width=2.0in]{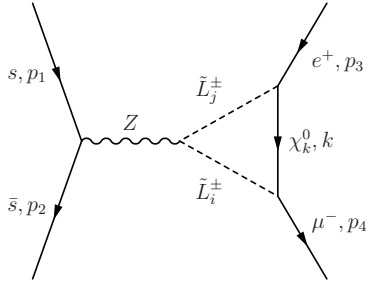}
\caption[]{Take Fig.\ref{fig1}(a) for example,and particle momentum is fixed.}
\label{fig3}
\end{center}
\end{figure}

Taking the diagram in Fig.\ref{fig1} as an example, we show how to write
the corresponding correction to the LFV decay $\phi\rightarrow e^+\mu^-$ in MSSM with seesaw mechanism.
At quark level, the relevant amplitude is written as:
\begin{eqnarray}
&&{\cal A}_Q=\frac{-e^{2}g_{\mu \nu }}{4s_{\rm w}^{2}c_{\rm w}^{2}}\int \frac{d^{D}k}{(2\pi )^{D}}
\frac{(Z_{L}^{2i}Z_{L}^{1j\ast}-2s_{W}^{2}\delta^{ij})(p_3+p_4)^{\nu}}{[(p_1+p_2)^{2}-m_{Z}^{2}]
[k^{2}-m_{\chi^{0}}^{2}]}
\nonumber\\
&&\hspace{1.0cm}\times
\frac{\bar{\upsilon}_{s}(p_2)\gamma^{\mu}(P_{L}-\frac{2}{3}s_{\rm w}^{2} )u_{s}(p_1)}{[(p_3-k)^{2}-m_{\tilde{L}}^{2}][(p_4+k)^{2}-m_{\tilde{L}}^{2}]}\bar{u}_{\mu}(p_4)
\nonumber\\
&&\hspace{1.0cm}\times
\Big\{\Big[Z_{L}^{2i\ast}(Z_{N}^{1k\ast}s_{W}+Z_{N}^{2k\ast}c_{W})
+Y_{l}^{2}Z_{L}^{5i\ast}Z_{N}^{3k\ast}\Big]P_{R}
\nonumber\\
&&\hspace{1.0cm}+\Big(Z_{L}^{5i\ast}Z_{N}^{1k}
+Y_{l}^{2}Z_{L}^{2i\ast}Z_{N}^{3k}\Big)P_{L}\Big\}\Big(\not{k}+m_{\chi^{0}_k}\Big)
\nonumber\\
&&\hspace{1.0cm}\times\Big\{\Big[Z_{L}^{1j}(Z_{N}^{1k}s_{W}+Z_{N}^{2k}c_{W})+Y_{l}^{1}Z_{L}^{4j}Z_{N}^{3k}\Big]P_{L}
\nonumber\\
&&\hspace{1.0cm}
+\Big(Z_{L}^{4j}Z_{N}^{1k\ast}+Y_{l}^{1}Z_{L}^{1j}Z_{N}^{3k\ast}\Big)P_{R}\Big\}\upsilon_{e}(p_3).
\label{eq3}
\end{eqnarray}
In the frame of center of mass, one can write down the amplitude at hadron level using
Eq.(\ref{hadron}):
\begin{eqnarray}
&&{\cal A}_H=\frac{-e^{2}}{24N_{c}s_{\rm w}^{2}c_{\rm w}^{2}}\int \frac{d^{D}k}{(2\pi )^{D}}
\frac{(Z_{L}^{2i}Z_{L}^{1j\ast}
-2s_{\rm w}^{2}\delta^{ij})}
{[(p_1+p_2)^{2}-m_{Z}^{2}][k^{2}-m_{\chi^{0}}^{2}]}
\nonumber\\
&&\hspace{1.0cm}\times
\frac{f_{\phi }m_{\phi }(4s_{\rm w}^{2}-3)\;\varepsilon(p)\cdot(p_3+p_4)}
{[(p_3-k)^{2}-m_{\tilde{L}}^{2}][(p_4+k)^{2}-m_{\tilde{L}}^{2}]}\bar{u}_{\mu}(p_4)
\nonumber\\
&&\hspace{1.0cm}\times
\Big\{\Big[Z_{L}^{2i\ast}(Z_{N}^{1k\ast}s_{W}+Z_{N}^{2k\ast}c_{W})
+Y_{l}^{2}Z_{L}^{5i\ast}Z_{N}^{3k\ast}\Big]P_{R}
\nonumber\\
&&\hspace{1.0cm}+\Big(Z_{L}^{5i\ast}Z_{N}^{1k}
+Y_{l}^{2}Z_{L}^{2i\ast}Z_{N}^{3k}\Big)P_{L}\Big\}\Big(\not{k}+m_{\chi^{0}_k}\Big)
\nonumber\\
&&\hspace{1.0cm}\times\Big\{\Big[Z_{L}^{1j}(Z_{N}^{1k}s_{W}+Z_{N}^{2k}c_{W})+Y_{l}^{1}Z_{L}^{4j}Z_{N}^{3k}\Big]P_{L}
\nonumber\\
&&\hspace{1.0cm}
+\Big(Z_{L}^{4j}Z_{N}^{1k\ast}+Y_{l}^{1}Z_{L}^{1j}Z_{N}^{3k\ast}\Big)P_{R}\Big\}\upsilon_{e}(p_3).
\label{eq5}
\end{eqnarray}
Applying the high energy physics package FeynCalc \cite{Mertig}, one can simplify
the amplitude in terms of invariant Passarino-Veltman integrals \cite{Passarino}:
\begin{eqnarray}
{\cal A}_H=&&\frac{i e^2\pi ^2f_{\phi }m_{\phi} (4 s_{\rm w}^{2}-3)}
{24 N_c s_{\rm w}^{2}c_{\rm w}^{2} (m_{\phi }^2-m_z^2) }
\sum_{i,j,k=1}^{6,6,4}(p_3+p_4)\cdot \varepsilon(p)
\nonumber\\
&&\times A_{5}^{ij}\bar{u}_\mu(p_4)
\Big\{C_1 m_e (A_{3}^{kj}A_{1}^{ik*}P_L +A_{4}^{kj}A_{2}^{ik*}P_R )
\nonumber\\
&&+C_2\Big[(m_eA_{3}^{kj}A_{1}^{ik*}-m_{\mu}A_{4}^{kj}A_{2}^{ik*})P_L
\nonumber\\
&&+(m_eA_{4}^{kj}A_{2}^{ik*}-m_{\mu}A_{3}^{kj}A_{1}^{ik*})P_R\Big]
\nonumber\\
&&+C_0\Big[(m_eA_{4}^{kj}A_{2}^{ik*}
-m_{\chi_k^0}A_{4}^{kj}A_{1}^{ik*}) P_R
\nonumber\\
&&+(m_eA_{3}^{kj}A_{1}^{ik*}-m_{\chi_k^0}A_{3}^{kj}A_{2}^{ik*})P_L\Big]\Big\}\upsilon_e(p_3)
\end{eqnarray}
All of integrals can be calculated through another high energy physics package LoopTools \cite{Hahn}.
In a similar way, we can write down the corrections from other diagrams in Fig.\ref{fig1} and Fig.\ref{fig2}
at hadron level, and list the simplified amplitudes in Appendix A.

Using the summation formula
\begin{eqnarray}
\sum_{\lambda =\pm 1,0}\varepsilon ^{\mu }_{\lambda }(p)\varepsilon ^{\ast \nu }_{\lambda }(p)
\equiv -g^{\mu \nu }+\frac{p^{\mu }p^{\nu }}{m_{\phi}^{2}},
\end{eqnarray}
we express the branching ratio of $\phi\rightarrow e^{+}\mu^{-}$ as
\begin{eqnarray}
Br(\phi\rightarrow e^+\mu^-)&=&\frac{\sqrt{[m_{\phi }^{2}-(m_{e}+m_{\mu})^{2}]
[m_{\phi}^{2}-(m_{e}-m_{\mu})^{2}]}}{16 \pi m_{\phi}^{3}\Gamma_{\phi}}
\nonumber\\
&&\times\sum_{i}^{}{\cal A}_{i}{\cal A}_{i}^{\ast},
\label{eq6}
\end{eqnarray}
in which $\Gamma_{\phi}$ is the total decay width, ${\cal A}_{i}$ are the amplitudes in Appendix A. The branching ratios for
$J/\Psi(\Upsilon)\rightarrow \mu^+\tau^-$ can be formulated in a similar way.

\section{Numerical Analysis}
\label{sec:3}
In the numerical analysis, we adopt the following value for mass of mesons $M_{\phi}=1.019$GeV, $M_{J/\Psi}=3.096$GeV, $M_{\Upsilon}=9.460$GeV. For the decay constants, we take $f_{\phi}=0.231$GeV, $f_{J/\Psi}=0.405$GeV, $f_{\Upsilon}=0.715$GeV \cite{Decay}. Furthermore, the electromagnetic coupling is determined by $\alpha(m_{_Z})=1/127$. Coinciding with the neutrino oscillation data and not losing generality, we always assume the lightest neutrino mass as: $m_{\nu^e_{L} }=1.0\times10^{-14}$GeV,
and the masses of three neutrinos satisfy following relations from experiment: $\Delta m_{sol}^{2}=8.0\times10^{-5} \mathrm{eV^{2}}$, $\Delta m_{atm}^{2}=3.0\times10^{-3} \mathrm{eV^{2}}$. Here, we also assume three right handed neutrinos are degenerate, i.e., $M_{\nu^e_R}\sim M_{\nu^{\mu}_R}\sim M_{\nu^{\tau}_R}\sim M_0$, $M_0$ is the mass scale of three right handed neutrinos. The recent results of the LHC experiments indicate that the lower limit of the squark mass is roughly given as 800 GeV\cite{LHC}. Not losing generality, we assume the degenerate spectrum in scalar quark sector $(m_{Q}^{2})_{IJ}=(m_{U}^{2})_{IJ}=(m_{D}^{2})_{IJ}=m_{\tilde{Q}}^{2}\delta_{IJ} =1{\rm TeV}^2,\;A^{IJ}_{q}=0$ ($I,\;J=1,\;2,\;3$) at GUT scale to satisfy the constraint. Through the calculation of mass spectrum and mixing matrices, a publicly available fortran77 program $SUSY\_FLAVOR$ is used \cite{Rosiek3}.

In our numerical analysis, we assume that the gaugino masses are GUT-related, that is,
\begin{eqnarray}
M_{1}= \frac{5s_{\rm w}^{2}}{3c_{\rm w}^{2}}M_{2},\;
M_{2}= \frac{\alpha _{2}}{\alpha _{s}}M_{3}\approx\frac{1}{3}M_{3}.
\end{eqnarray}
In order to decrease the number of free parameters involved in our calculation,
we suppose that the diagonal entries of two $3\times3$ matrices $m_{L}^{2}$, $m_{R}^{2}$
in Eq.(\ref{eq2}) are equal $(m_{L}^{2})_{II}=(m_{R}^{2})_{II}=m_{\tilde{E}}^{2}$, where
$I=1,\;2,\;3$. Now, the only sources of LFV are off-diagonal entries of the soft breaking terms
$m_{L}^{2}$, $m_{R}^{2}$ and $A_{l}$. Those off-diagonal entries of $3\times3$ matrices
$m_{L}^{2}$, $m_{R}^{2}$ are parameterized by mass insertions as in \cite{Rosiek3},
\begin{eqnarray}
\Big(m^{2}_{L}\Big)^{IJ}&=&\delta ^{IJ}_{L}\sqrt{(m^{2}_{L})^{II}(m^{2}_{L})^{JJ}},\\
\Big(m^{2}_{R}\Big)^{IJ}&=&\delta ^{IJ}_{R}\sqrt{(m^{2}_{R})^{II}(m^{2}_{R})^{JJ}}.
\end{eqnarray}
where $I,\;J=1,\;2,\;3$. We also assume $\delta ^{IJ}_{L}=\delta ^{IJ}_{R}$. Meanwhile, the trilinear soft breaking coupling is parameterized by
\begin{eqnarray}
A^{II}_{l}=a^{I}_{l}Y^{I}_{l}\sqrt[4]{(m_{L}^{2})^{II}(m_{R}^{2})^{JJ}},\\
A^{IJ}_{l}=\delta^{IJ}_{LR}\sqrt{2(m_{L}^{2})^{II}(m_{R}^{2})^{JJ}}.
\end{eqnarray}
At first, we discuss the LFV decays of vector mesons $\phi\rightarrow e^+\mu^-$. The corrections from Higgs to the LFV branching ratios of vector mesons $\phi\rightarrow e^+\mu^-$ can be neglected safely since they are suppressed by the tiny masses of leptons.

In the MSSM with type I seesaw, the LFV processes originate from the mass insertions $\delta_{L}^{ij},\;\delta_{R}^{ij}$. The most challenging experimental prospects arise for the CR($\mu-e$) in heavy nuclei such as titanium ($_{22}^{48}\textrm{Ti}$). The experimental upper bounds on the conversion rate reach CR $(\mu-e, Ti)\le 4.3\times10^{-12}$ \cite{PDG}. In the MSSM with type I seesaw, the conversion rate in nuclei can be calculated by \cite{SeeSaw1}:
\begin{eqnarray}
CR\Big(\mu-e,X\Big)&=&{\Gamma(\mu+X\rightarrow e+X)\over\Gamma(\mu+X\rightarrow capture)}\nonumber\\
&=&4\alpha^5{Z_{eff}^4\over Z}|F(q)|^2 m_{\mu}^5\Big[\Big|Z(A_1^L-A_2^R)\nonumber\\
&&-(2Z+N)\bar{D}_u^L-(Z+2N)\bar{D}_d^L\Big|^2\nonumber\\
&&+\Big|Z(A_1^R-A_2^L)-(2Z+N)\bar{D}_u^R\nonumber\\
&&-(Z+2N)\bar{D}_d^R\Big|^2\Big],
\end{eqnarray}
where Z and N denote the proton and neutron numbers in a nucleus,
$F(q^2)$ is the nuclear form factor and $Z_{eff}$ is an effective atomic charge, $A_{1,2}^{L,R}$ stand for the contributions from penguin-type diagram, $\bar{D}_{u,d}^{L,R}$ stand for the contributions from box-
type diagrams. In $_{22}^{48}\textrm{Ti}$, $F(q^2)\sim0.54$ and $Z_{eff}=17.6$ \cite{Kitano}. After a scan over the parameter space, we will discuss $\phi\rightarrow e^+\mu^-$ and $\mu-e$ conversion for two cases: (I) $\delta^{12}_{L}$ dominance, $\delta^{23}_{L} = \delta^{13}_{L} = 0$; (II) $\delta^{23}_{L}\delta^{13}_{L}$ dominance, $\delta^{12}_L = 0$. We assume $tan\beta=10$, $\mu=200$GeV, $M_{2}=200$GeV, $a^{I}_{l}=1$ and $\delta^{IJ}_{LR}=0$.

\begin{figure}
\begin{center}
\includegraphics[width=0.6\columnwidth]{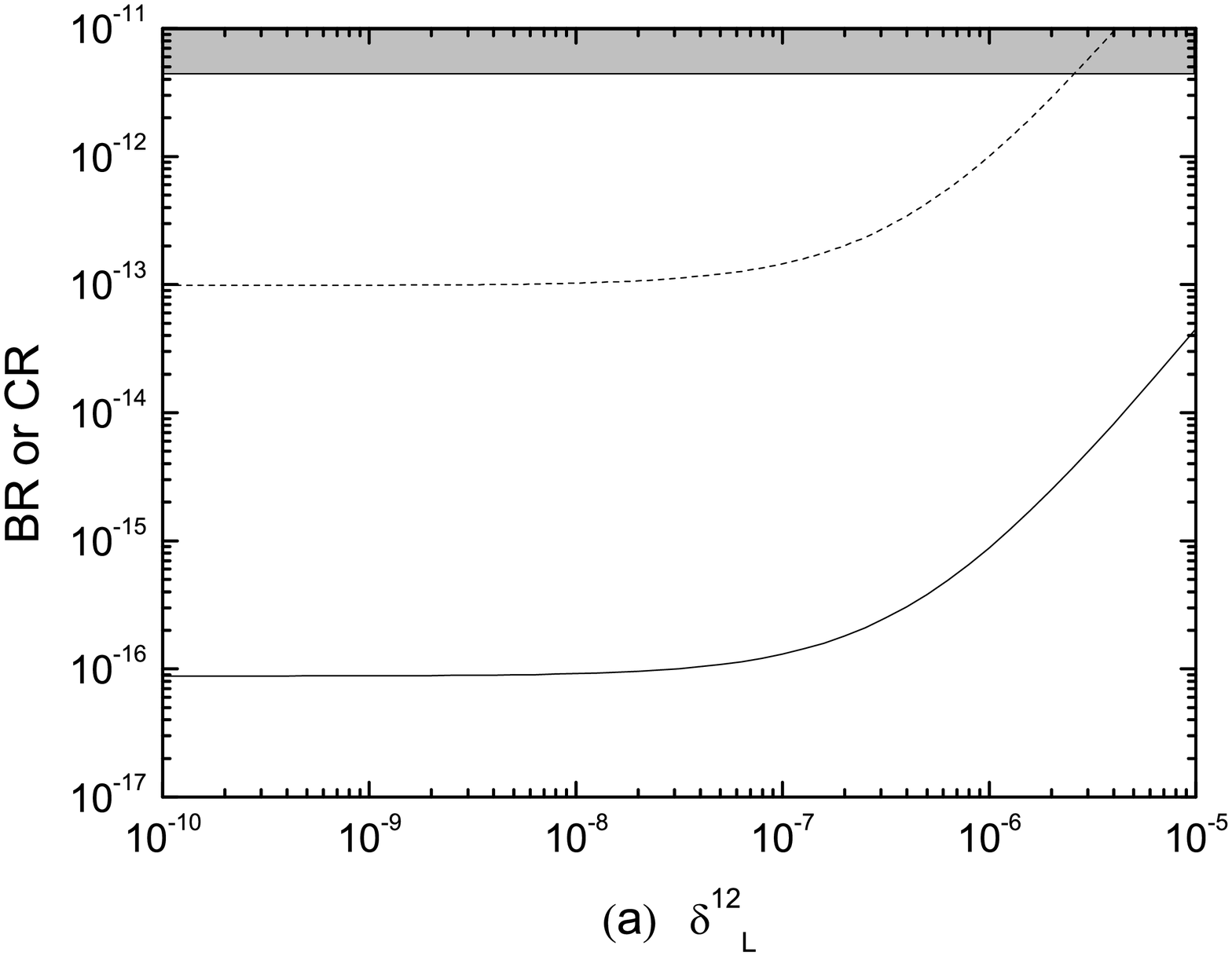}
\includegraphics[width=0.6\columnwidth]{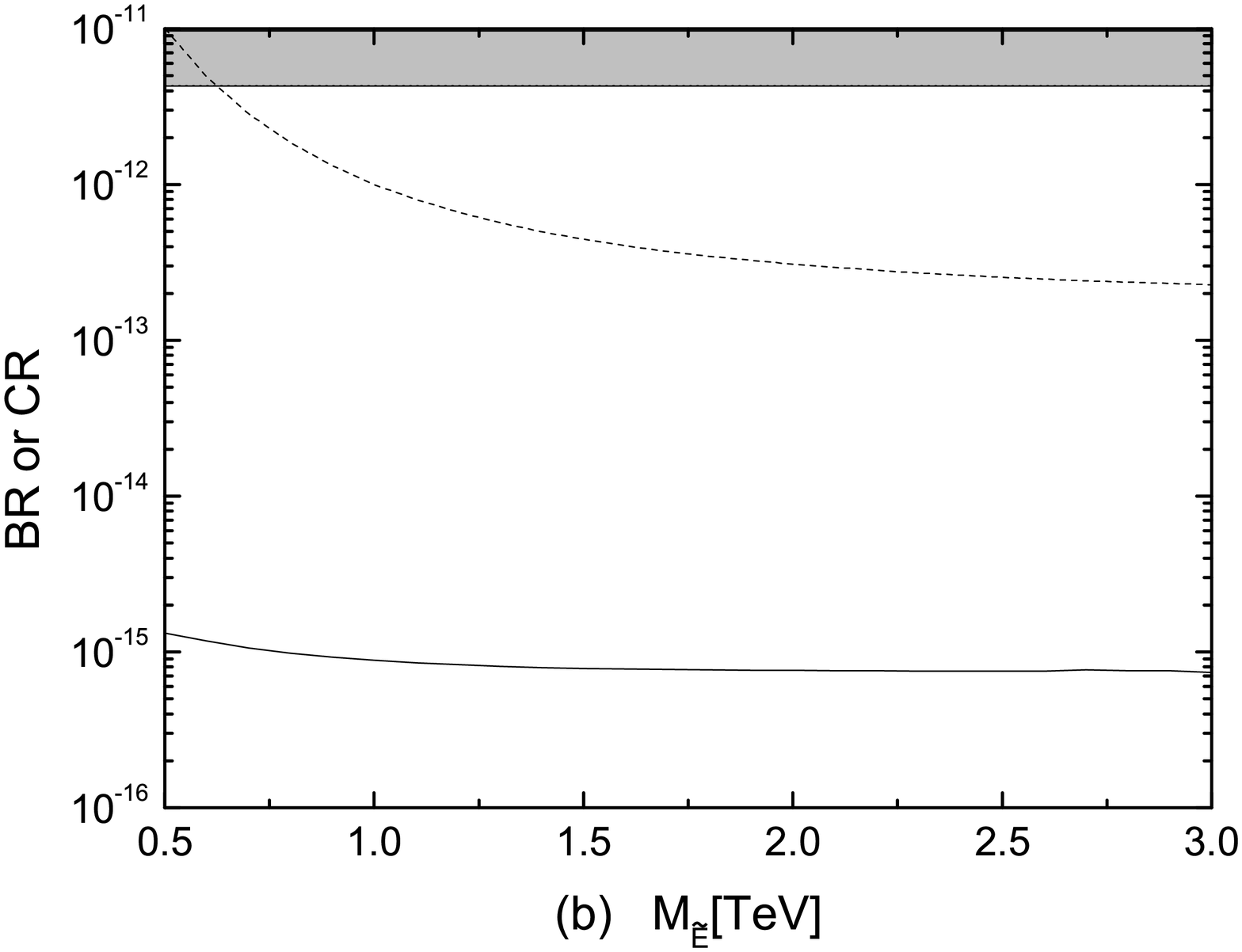}
\includegraphics[width=0.6\columnwidth]{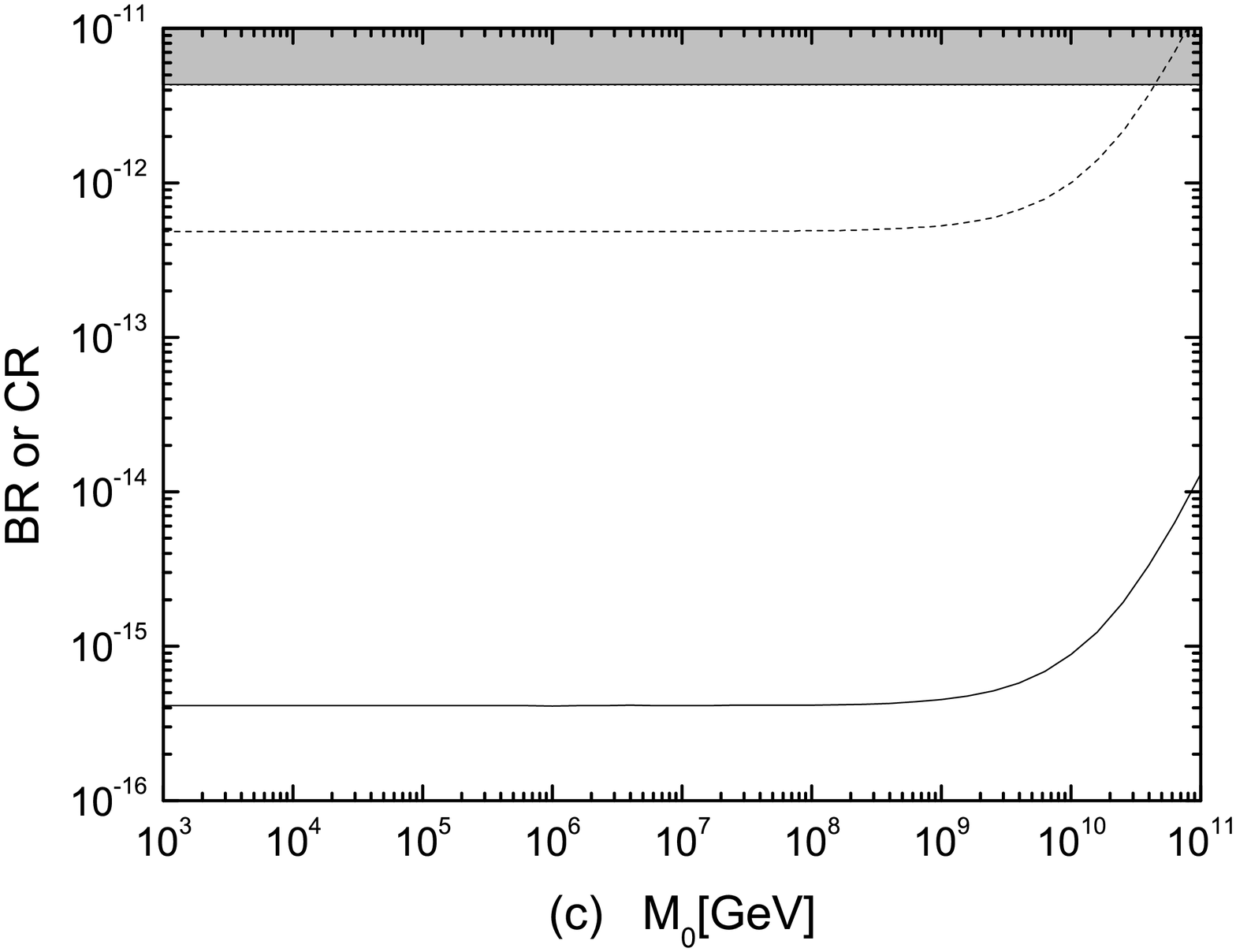}
\caption[]{Case (I):BR ($\phi\rightarrow e^+\mu^-$) (solid line) and CR$(\mu+_{22}^{48}\textrm{Ti}\rightarrow e+_{22}^{48}\textrm{Ti})$ (dash line) vs mass insertion $\delta^{12}_{L}$, slepton mass sector $m_{\tilde E}$ and right handed neutrinos mass scale $M_{0}$. The shadow is the excluded region for CR$(\mu+_{22}^{48}\textrm{Ti}\rightarrow e+_{22}^{48}\textrm{Ti})$.}
\label{fig4}
\end{center}
\end{figure}
\begin{figure}
\begin{center}
\includegraphics[width=0.6\columnwidth]{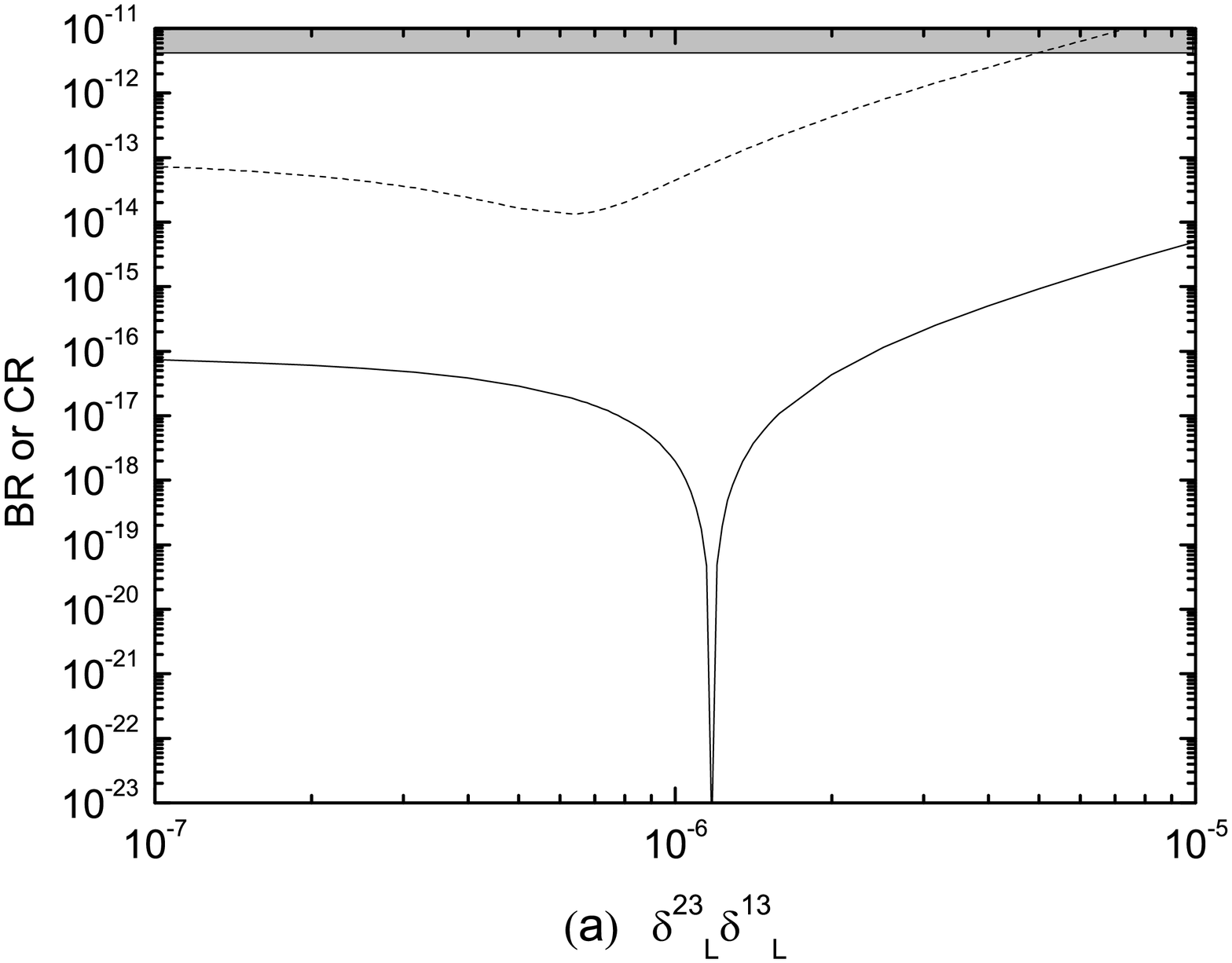}
\includegraphics[width=0.6\columnwidth]{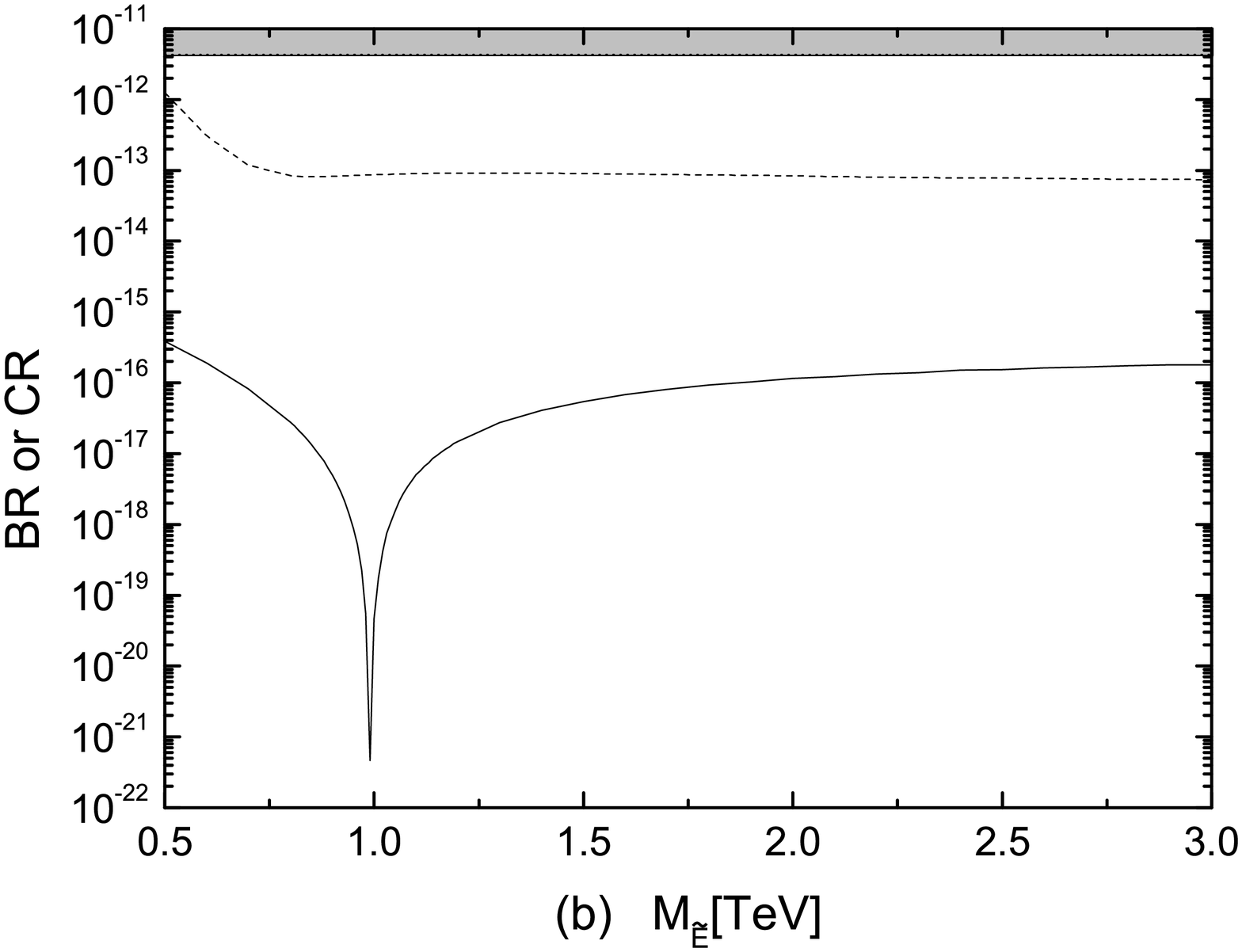}
\includegraphics[width=0.6\columnwidth]{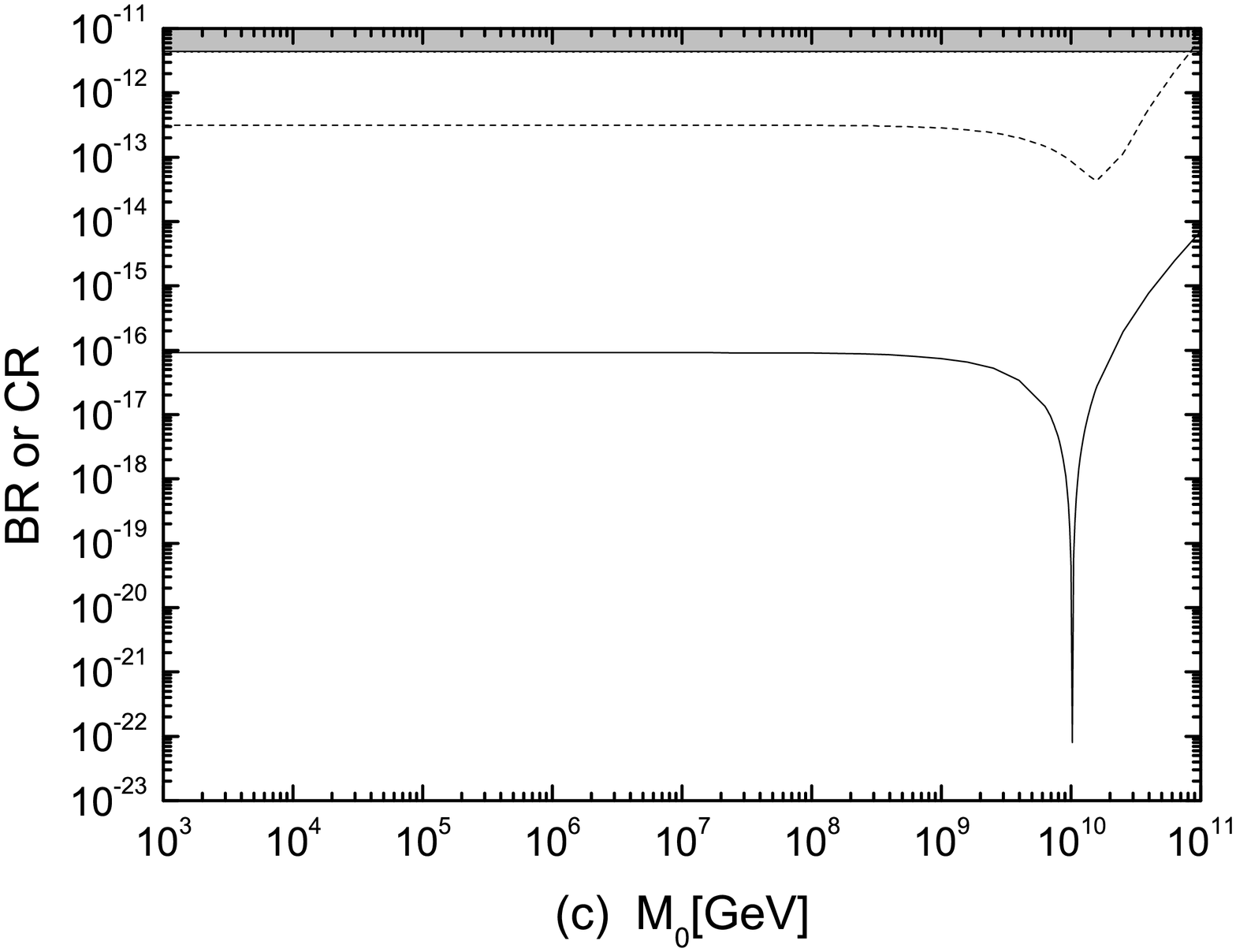}
\caption[]{Case (II):BR ($\phi\rightarrow e^+\mu^-$) (solid line) and CR$(\mu+_{22}^{48}\textrm{Ti}\rightarrow e+_{22}^{48}\textrm{Ti})$ (dash line) vs mass insertion $\delta^{23}_{L}\delta^{13}_{L}$, slepton mass sector $m_{\tilde E}$ and right handed neutrinos mass scale $M_{0}$. The shadow is the excluded region for CR$(\mu+_{22}^{48}\textrm{Ti}\rightarrow e+_{22}^{48}\textrm{Ti})$.}
\label{fig5}
\end{center}
\end{figure}
{\bf Case (I)} Taking $m_{\tilde E}=1$TeV, $M_{0}=10^{10}$GeV, we plot the theoretical prediction of BR $(\phi\rightarrow e^+\mu^-)$ (solid line) and CR$(\mu+_{22}^{48}\textrm{Ti}\rightarrow e+_{22}^{48}\textrm{Ti})$ (dash line) versus $\delta^{12}_{L}$ in Fig.\ref{fig4}(a), where the gray shadow is the excluded region for CR$(\mu+_{22}^{48}\textrm{Ti}\rightarrow e+_{22}^{48}\textrm{Ti})$ . The CR$(\mu+_{22}^{48}\textrm{Ti}\rightarrow e+_{22}^{48}\textrm{Ti})$ exceeds the current experiment limit at $\delta^{12}_{L}\sim 2.0\times 10^{-6}$. The parameter space of $\delta^{12}_{L}$ has been highly suppressed with respect to the prediction of CR$(\mu+_{22}^{48}\textrm{Ti}\rightarrow e+_{22}^{48}\textrm{Ti})$. Both BR $(\phi\rightarrow e^+\mu^-)$ and CR$(\mu+_{22}^{48}\textrm{Ti}\rightarrow e+_{22}^{48}\textrm{Ti})$ tend to be not sensitive to $\delta^{12}_{L}$ when its value is below $10^{-7}$. In \cite{SeeSaw1}, the authors investigate the LFV processes $\mu\rightarrow e\gamma$, $\mu\rightarrow 3e$ and deduce a constraint with $\delta^{12}_{L}\le 3\times10^{-4}$. Comparing with the constraint on $\delta^{12}_{L}$ from $\mu\rightarrow e\gamma$, $\mu\rightarrow 3e$, It displays the constraint from $\mu-e$ conversion is more stronger.

Taking $M_{0}=10^{10}$GeV, $\delta^{12}_{L}=10^{-6}$, we plot the theoretical prediction of BR $(\phi\rightarrow e^+\mu^-)$ (solid line) and CR$(\mu+_{22}^{48}\textrm{Ti}\rightarrow e+_{22}^{48}\textrm{Ti})$ (dash line) versus slepton mass sector $m_{\tilde E}$ in Fig.\ref{fig4}(b), where the gray shadow is the excluded region for CR$(\mu+_{22}^{48}\textrm{Ti}\rightarrow e+_{22}^{48}\textrm{Ti})$ . Different to Fig.\ref{fig4}(a) and Fig.\ref{fig4}(c), Both BR $(\phi\rightarrow e^+\mu^-)$ and CR$(\mu+_{22}^{48}\textrm{Ti}\rightarrow e+_{22}^{48}\textrm{Ti})$ decrease as $m_{\tilde E}$ varies from 0.5TeV to 3TeV. For lower slepton mass, the prediction on CR$(\mu+_{22}^{48}\textrm{Ti}\rightarrow e+_{22}^{48}\textrm{Ti})$ is also out of the current experiment limit.

Taking $m_{\tilde E}=1$TeV, $\delta^{12}_{L}=10^{-6}$, we plot the theoretical prediction of BR $(\phi\rightarrow e^+\mu^-)$ (solid line) and CR$(\mu+_{22}^{48}\textrm{Ti}\rightarrow e+_{22}^{48}\textrm{Ti})$ (dash line) versus the right handed neutrino mass scale $M_{0}$ in Fig.\ref{fig4}(c), where the gray shadow is the excluded region for CR$(\mu+_{22}^{48}\textrm{Ti}\rightarrow e+_{22}^{48}\textrm{Ti})$. Both BR $(\phi\rightarrow e^+\mu^-)$ and CR$(\mu+_{22}^{48}\textrm{Ti}\rightarrow e+_{22}^{48}\textrm{Ti})$
also show a strong dependence on  $M_{0}$ in range of $M_{0}\ge10^{10}$GeV,
but most part is ruled out by the constraint from $\mu - e$ conversion.
When $M_{0}\le10^{10}$GeV, the dependence becomes weaker and weaker.

{\bf Case (II)} Taking $m_{\tilde E}=1$TeV, $M_{0}=10^{10}$GeV, we plot the theoretical prediction of BR $(\phi\rightarrow e^+\mu^-)$ (solid line) and CR$(\mu+_{22}^{48}\textrm{Ti}\rightarrow e+_{22}^{48}\textrm{Ti})$ (dash line) versus $\delta^{23}_{L}\delta^{13}_{L}$ in Fig.\ref{fig5}(a), where the gray shadow is the excluded region for CR$(\mu+_{22}^{48}\textrm{Ti}\rightarrow e+_{22}^{48}\textrm{Ti})$ . There is a sharp decrease around $\delta^{23}_{L}\delta^{13}_{L}\sim 1.17\times10^{-6}$ with a minimum BR $(\phi\rightarrow e^+\mu^-)$ of order about $10^{-23}$, which is about two orders smaller than the most stringent prediction in \cite{Gutsche}. In \cite{SeeSaw1}, the authors also give a expected value for $\delta^{23}_{L}\delta^{13}_{L}$ deduced from the processes $\mu\rightarrow e\gamma$ and $\mu\rightarrow 3e$, which is $(\sim10^{-6})$ and compatible with ours.

Taking $M_{0}=10^{10}$GeV, $\delta^{23}_{L}\delta^{13}_{L}=1.2\times10^{-6}$, we plot the theoretical prediction of BR $(\phi\rightarrow e^+\mu^-)$ (solid line) and CR$(\mu+_{22}^{48}\textrm{Ti}\rightarrow e+_{22}^{48}\textrm{Ti})$ (dash line) versus $m_{\tilde E}$ in Fig.\ref{fig5}(b), where the gray shadow is the excluded region for CR$(\mu+_{22}^{48}\textrm{Ti}\rightarrow e+_{22}^{48}\textrm{Ti})$. Here and followed, we will assume $\delta^{23}_{L}=4\times10^{-3}$, $\delta^{13}_{L}=3\times10^{-4}$ and these are also expected values for $\delta^{23}_{L}$ and $\delta^{13}_{L}$ reported in \cite{SeeSaw1} evaluated from LFV decays $\tau\rightarrow \mu\gamma$ and $\tau\rightarrow e\gamma$. We also find a resonating absorption around $m_{\tilde E}=1$TeV that originates from the interference between the corrections from sneutrino sector and that from charged slepton sector. Fig. 5(b) displays that no constraint on $m_{\tilde E}$ has arisen with respect to $\mu-e$ conversion and LFV decay $\phi\rightarrow e^+\mu^-$.

Taking $m_{\tilde E}=1$TeV, $\delta^{23}_{L}\delta^{13}_{L}=1.2\times10^{-6}$, we plot the theoretical prediction of BR $(\phi\rightarrow e^+\mu^-)$ (solid line) and CR$(\mu+_{22}^{48}\textrm{Ti}\rightarrow e+_{22}^{48}\textrm{Ti})$ (dash line) versus $M_{0}$ in Fig.\ref{fig5}(c), where the gray shadow is the excluded region for CR$(\mu+_{22}^{48}\textrm{Ti}\rightarrow e+_{22}^{48}\textrm{Ti})$.There is a resonating absorption around $M_{0}=10^{10}$GeV that originates from the interference between the corrections from sneutrino sector.

Comparing Case (I) with Case (II), we find: (i) In Case (II), the prediction for BR $(\phi\rightarrow e^+\mu^-)$ are more compatible with \cite{Gutsche}. In \cite{zhang} and \cite{Gutsche}, the constraints are BR $(\phi\rightarrow e^+\mu^-)\le4.0\times 10^{-17}$ and BR $(\phi\rightarrow e^+\mu^-)\le1.3\times 10^{-21}$. It is noted worthwhile that the prediction in \cite{zhang} also satisfies the constraint from $\mu-e$ conversion, even if it is derived by the constraint from $\mu\rightarrow3e$. (ii) Compared with $\delta^{23}_L$ ($\sim 10^{-3}$) and $\delta^{13}_L$ ($\sim 10^{-4}$), the value for $\delta^{12}_L$ ($10^{-6}$ or little) is so small that it can be neglected.

\begin{figure}
\begin{center}
\includegraphics[width=0.6\columnwidth]{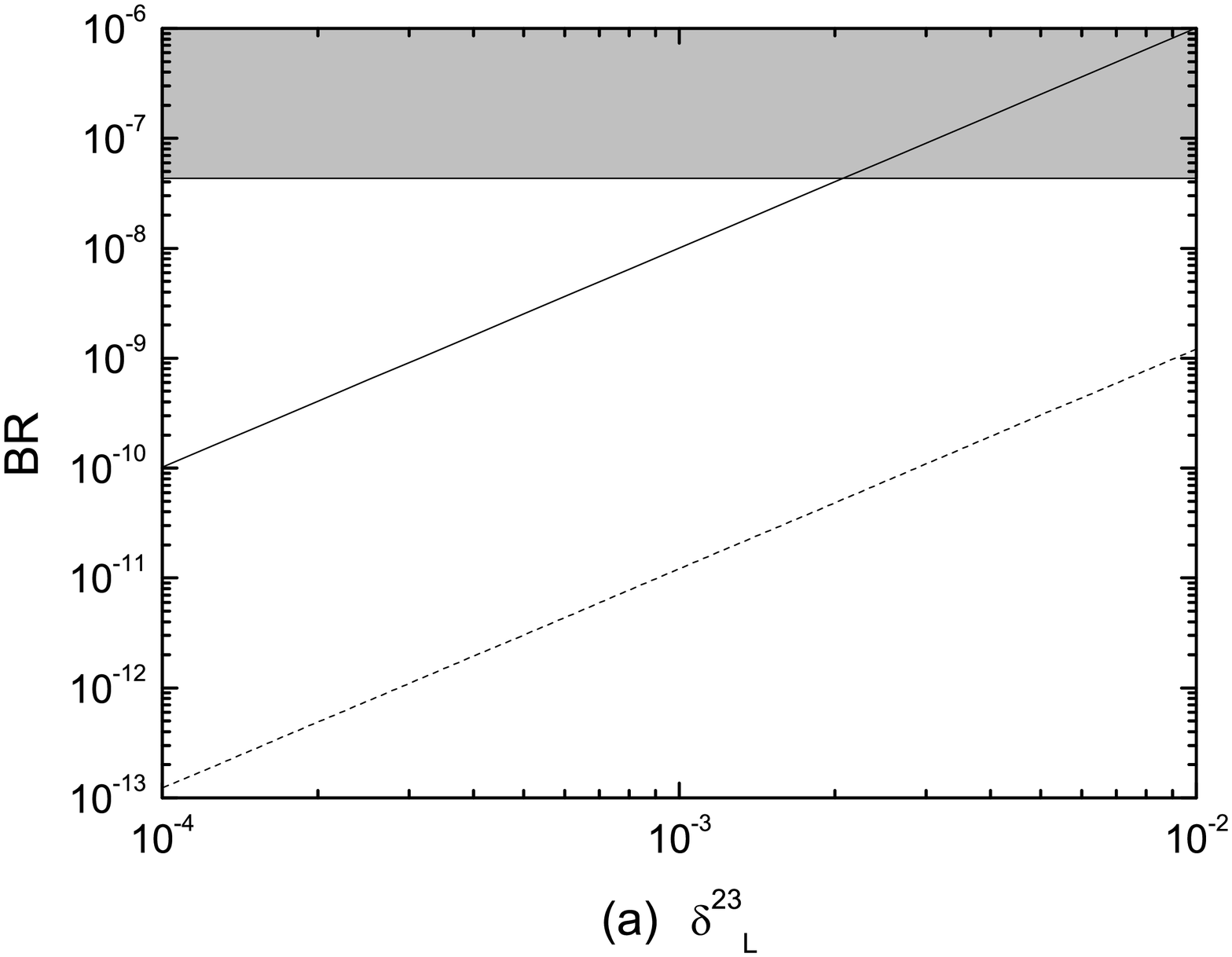}
\includegraphics[width=0.6\columnwidth]{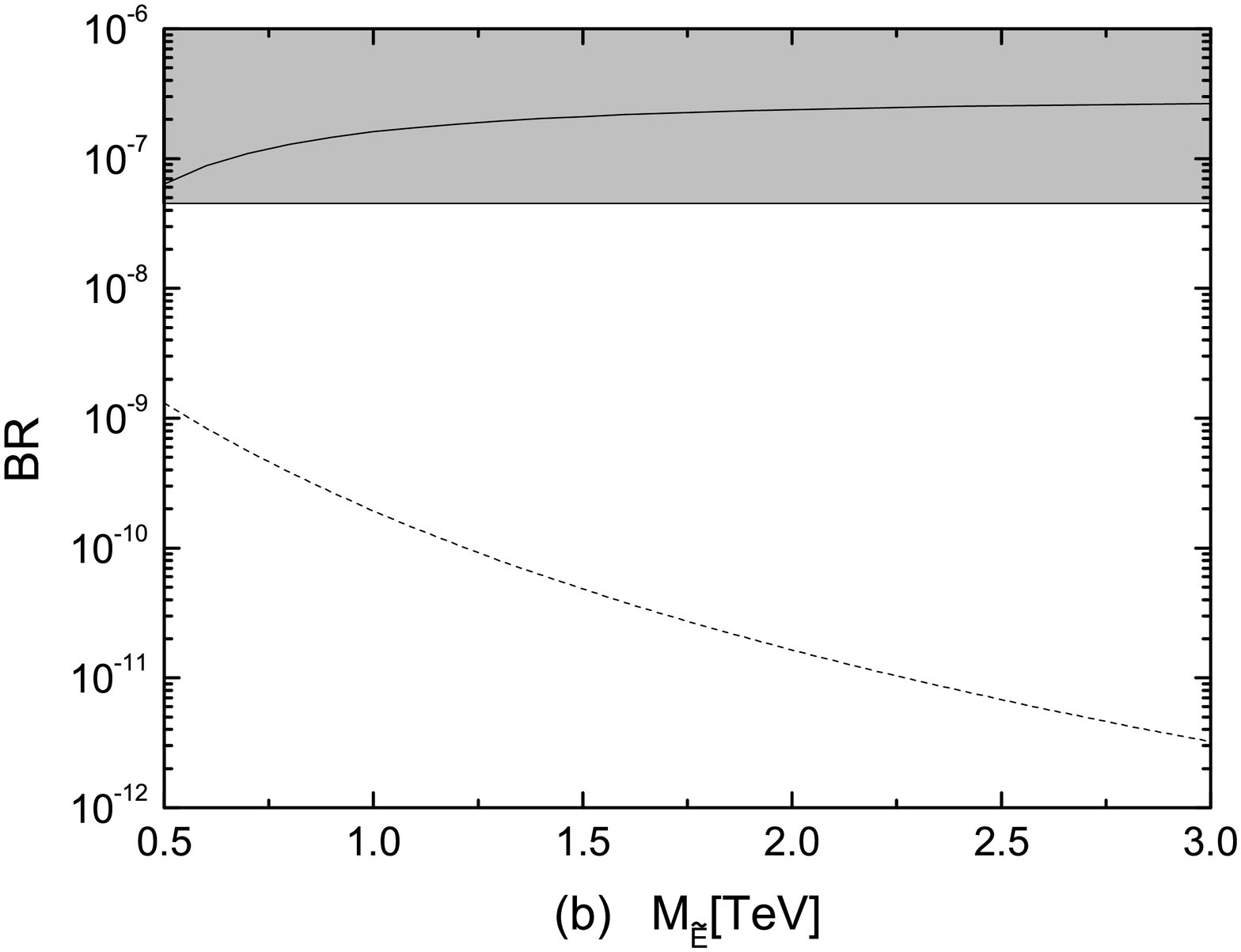}
\caption[]{BR ($J/\Psi\rightarrow \mu^+\tau^-$) (solid line) and BR$(\tau\rightarrow \mu\gamma)$ (dash line) vs mass insertion $\delta^{23}_{L}$, slepton sector $m_{\tilde E}$. The gray shadow is the excluded region for BR$(\tau\rightarrow \mu\gamma)$.}
\label{fig6}
\end{center}
\end{figure}

\begin{figure}
\begin{center}
\includegraphics[width=0.6\columnwidth]{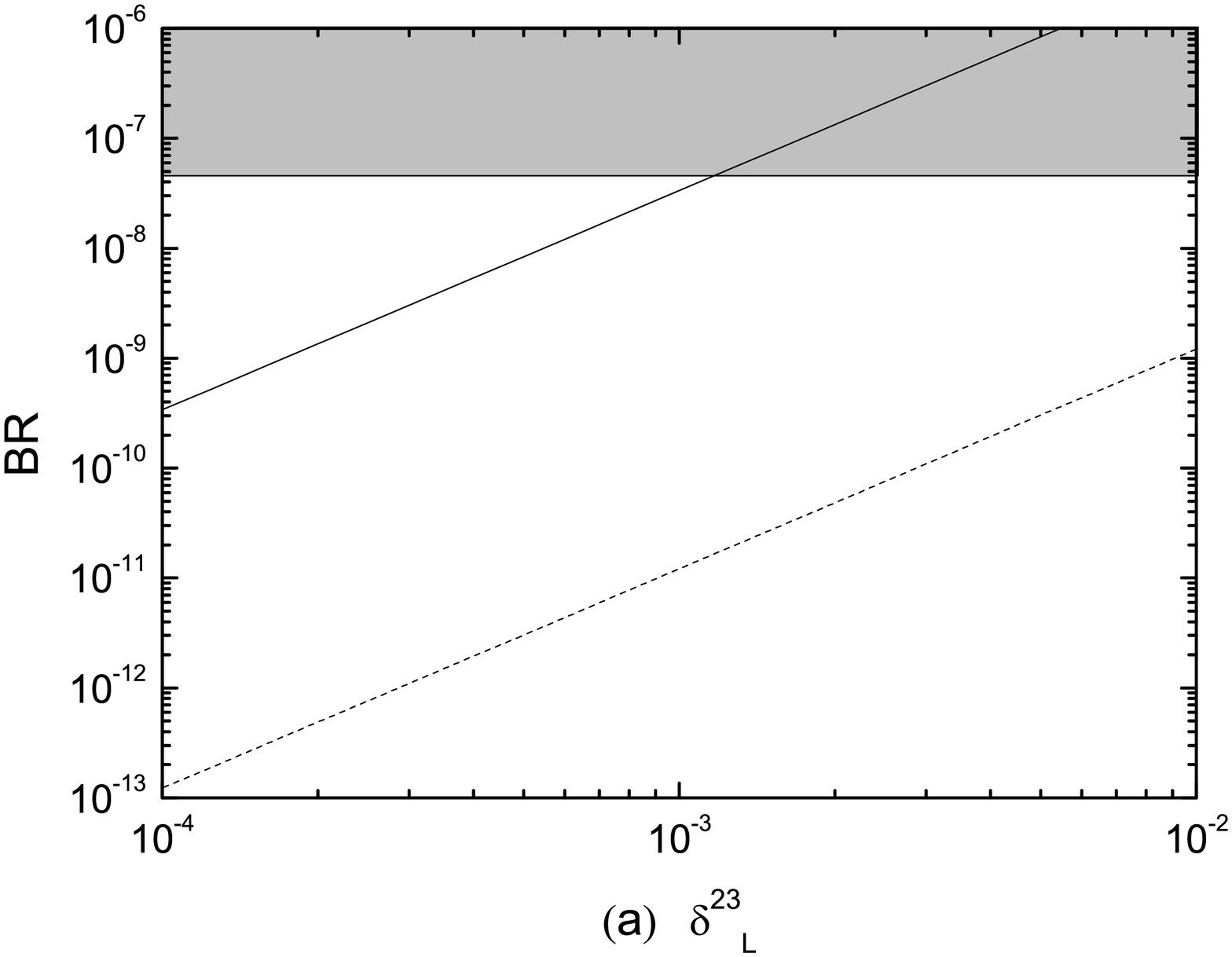}
\includegraphics[width=0.6\columnwidth]{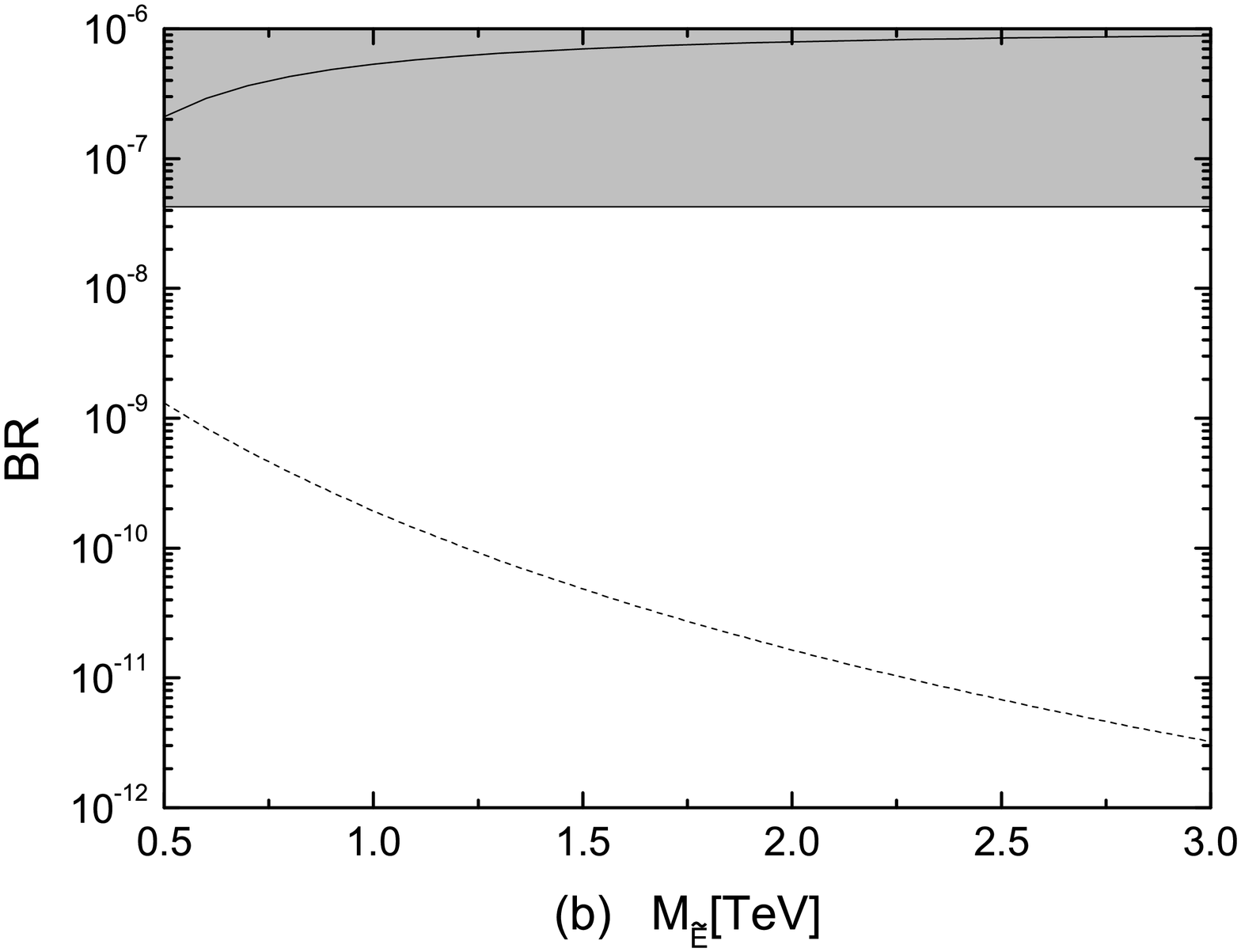}
\caption[]{BR($\Upsilon \rightarrow \mu^+\tau^-$) (solid line) and BR$(\tau\rightarrow \mu\gamma)$ (dash line) vs mass insertion $\delta^{23}_{L}$, slepton sector $m_{\tilde E}$. The gray shadow is the excluded region for BR$(\tau\rightarrow \mu\gamma)$.}
\label{fig7}
\end{center}
\end{figure}

Then, we will investigate meson decays $J/\Psi(\Upsilon)\rightarrow \mu^+\tau^-$ in Case (II) not only for reasons above, but also for the aim to generate a large enough BR $(J/\Psi(\Upsilon)\rightarrow \mu^+\tau^-)$ to be observed in experiment. As it displays in Fig.\ref{fig6} and Fig.\ref{fig7}, the mass insertion $\delta_L^{23}$ affects the theoretical evaluation of BR $(J/\Psi(\Upsilon)\rightarrow \mu^+\tau^-)$ strongly. In formula, there is a simple relation \cite{SeeSaw1}:
\begin{eqnarray}
\frac{BR(\tau\rightarrow3\mu)}{BR(\tau\rightarrow\mu\gamma)}\simeq\frac{\alpha}{8\pi}
\Big(\frac{16}{3}ln\frac{m_\tau}{2m_\mu}-\frac{14}{9}\Big)\simeq0.003.
\end{eqnarray}
So, we just consider the constraint from $\tau\rightarrow\mu\gamma$.

{\bf(I)$J/\Psi\rightarrow \mu^+\tau^-$}

Taking $m_{\tilde E}=1$TeV, $M_{0}=10^{10}$GeV, we plot the theoretical prediction of BR $(J/\Psi\rightarrow \mu^+\tau^-)$ (solid line) and BR$(\tau\rightarrow \mu\gamma)$ (dash line) versus $\delta^{23}_{L}$ in Fig.\ref{fig6}(a). Taking $M_{0}=10^{10}$GeV, $\delta^{23}_{L}=4\times10^{-3}$, $\delta^{13}_{L}=3\times10^{-4}$, we plot the theoretical prediction of BR $(J/\Psi\rightarrow \mu^+\tau^-)$ (solid line) and BR$(\tau\rightarrow \mu\gamma)$ (dash line) versus $m_{\tilde E}$ in Fig.\ref{fig6}(b). The gray shadow is the excluded region for BR$(\tau\rightarrow \mu\gamma)$ from the experiment. A linear relationship is displayed between BR $(J/\Psi\rightarrow \mu^+\tau^-)$ and $\delta^{23}_{L}$ in logarithmic scale, which shows the great dependence of BR $(J/\Psi\rightarrow \mu^+\tau^-)$ on $\delta^{23}_{L}$. We also investigate the dependence of BR $(J/\Psi\rightarrow \mu^+\tau^-)$ on $M_{0}$ and it shows BR $(J/\Psi\rightarrow \mu^+\tau^-)$ is not sensitive to $M_{0}$.

In \cite{zhang} and \cite{Bordes}, the authors calculate the BR $(J/\Psi\rightarrow \mu^+\tau^-)$ with BR $(J/\Psi\rightarrow \mu^+\tau^-)\le4.1\times10^{-9}$ and BR $(J/\Psi\rightarrow \mu^+\tau^-)\le6.3\times10^{-10}$, that is three or four orders below the experiment limit $2.0\times10^{-6}$. Under our assumption, $m_{\tilde E}=1$TeV, $M_0=10^{10}$GeV, $\delta^{23}_{L}=4\times10^{-3}$, $\delta^{13}_{L}=3\times10^{-4}$, we get the BR $(J/\Psi\rightarrow \mu^+\tau^-)$ can be enhanced as large as $1.6\times10^{-7}$, that is more promising to detect directly in experiment in near future.

{\bf(II)$\Upsilon\rightarrow \mu^+\tau^-$}

Similar to the process $J/\Psi\rightarrow \mu^+\tau^-$, the LFV decay $\Upsilon\rightarrow \mu^+\tau^-$ has the same behavior as a function of $\delta^{23}_{L}$, $m_{\tilde E}$ and $M_{0}$. Taking $m_{\tilde E}=1$TeV, $M_{0}=10^{10}$GeV, we plot the theoretical prediction of BR $(\Upsilon\rightarrow \mu^+\tau^-)$ (solid line) and BR$(\tau\rightarrow \mu\gamma)$ (dash line) versus $\delta^{23}_{L}$ in Fig.\ref{fig7}(a). Taking $M_{0}=10^{10}$GeV, $\delta^{23}_{L}=4\times10^{-3}$, $\delta^{13}_{L}=3\times10^{-4}$, we plot the theoretical prediction of BR$(\Upsilon\rightarrow \mu^+\tau^-)$ (solid line) and BR$(\tau\rightarrow \mu\gamma)$ (dash line) versus $m_{\tilde E}$ in Fig.\ref{fig7}(b). The gray shadow is the excluded region for BR$(\tau\rightarrow \mu\gamma)$ from the experiment. Similarly, the mass insertion $\delta_L^{23}$ affects the theoretical evaluation of BR $(\Upsilon\rightarrow \mu^+\tau^-)$ strongly, and BR $(\Upsilon\rightarrow \mu^+\tau^-)$ depends on $M_{0}$ mildly.

The most stringent prediction of BR $(\Upsilon\rightarrow \mu^+\tau^-)$ in theory is given in \cite{Bordes} with BR $(\Upsilon\rightarrow \mu^+\tau^-)\le2.9\times10^{-6}$, and that is very close to the experiment limit. In \cite{zhang}, the author calculates the BR $(\Upsilon\rightarrow \mu^+\tau^-)\le 7.2\times10^{-5}$, and it shows the updated data for BR$(\tau\rightarrow 3\mu)$ from experiment is necessary. Under our assumption, $m_{\tilde E}=1$TeV, $M_0=10^{10}$GeV, $\delta^{23}_{L}=4\times10^{-3}$, $\delta^{13}_{L}=3\times10^{-4}$, we get BR $(\Upsilon\rightarrow \mu^+\tau^-)\le5.3\times10^{-7}$. That is also promising to detect directly in experiment in near future.

We can evaluate the branching ratios of LFV decays $\rho(\omega, J/\Psi, \Upsilon)\rightarrow e^+\mu^-$ using above method. To shorten the length of text, we just present the upper bounds on those branching ratios under the same assumptions as $\phi\rightarrow e^+\mu^-$. After considering the constraints from $\mu-e$ conversion, $\mu\rightarrow e\gamma$, $\mu\rightarrow 3e$ etc,
we give a summary of upper bounds of experiment data and corresponding theoretical predictions in Tab.\ref{tab1}.
\begin{table*}
\caption{The upper bounds on the branching ratios of vector bosons}
\label{tab1}
\begin{tabular*}{\textwidth}{@{\extracolsep{\fill}}lllll@{}}
\hline
Decay & \multicolumn{1}{c}{Experiment} & \multicolumn{1}{c}{Ref.\cite{zhang}} & \multicolumn{1}{c}{Ref.\cite{Gutsche}} & Our prediction \\
\hline
$\phi\rightarrow e^+\mu^-$&$\le2.0\times 10^{-6}$&$\le4.0\times10^{-17}$&$\le1.3\times10^{-21}$&$\le 5.0\times10^{-20}$\\
$\rho\rightarrow e^+\mu^-$&$-$&$\le3.8\times10^{-20}$&$\le3.5\times10^{-24}$&$\le 1.0\times10^{-20}$\\
$\omega\rightarrow e^+\mu^-$&$-$&$\le8.1\times10^{-16}$&$\le6.2\times10^{-27}$&$\le 1.8\times10^{-20}$\\
$J/\Psi\rightarrow e^+\mu^-$&$<1.1\times10^{-6}$&$\le4.0\times10^{-13}$&$\le3.5\times10^{-13}$&$\le 1.9\times10^{-18}$\\
$\Upsilon\rightarrow e^+\mu^-$&$-$&$\le2.0\times10^{-9}$&$\le3.9\times10^{-6}$&$\le 3.6\times10^{-18}$\\
$J/\Psi\rightarrow \mu^+\tau^-$&$<2.0\times10^{-6}$&$\le4.1\times10^{-9}$&$-$&$\le 1.6\times10^{-7}$\\
$\Upsilon\rightarrow \mu^+\tau^-$&$<6.0\times10^{-6}$&$\le7.2\times10^{-5}$&$-$&$\le 5.3\times10^{-7}$\\
\hline
\end{tabular*}
\end{table*}

\section{Conclusions\label{sec:4}}
Considering the constraints from $\mu-e$ conversion, $\mu\rightarrow e\gamma$, $\tau\rightarrow\mu\gamma$, $\mu\rightarrow3e$ etc, we analyze the LFV decays of $\phi\rightarrow e^+\mu^-$,$J/\Psi(\Upsilon)\rightarrow\mu^+\tau^-$ in the framework of MSSM with type I seesaw extended.

In the MSSM with type I seesaw extended, the theoretical evaluation on BR $(\phi\rightarrow e^+\mu^-)$ is
affected by the mass insertion $\delta_L^{23}\delta_L^{13}$. After considering the constraints from $\mu-e$ conversion, $\mu\rightarrow e\gamma$ and $\mu\rightarrow3e$, the prediction on BR $(\phi\rightarrow e^+\mu^-)$ can reach $5.0\times10^{-20}$, which is far below the present experimental upper bound. In a similar way, the mass insertion $\delta_L^{23}$ affects the theoretical evaluations on BR $(J/\Psi\rightarrow \mu^+\tau^-)$ and BR $(\Upsilon\rightarrow \mu^+\tau^-)$ sensitively. Considering the constraints from $\tau\rightarrow \mu\gamma$ and $\tau\rightarrow3\mu$, the predictions on BR $(J/\Psi\rightarrow \mu^+\tau^-)$ and BR $(\Upsilon\rightarrow \mu^+\tau^-)$ can reach $10^{-7}$, which are very promising to be observed in near future experiment.

In the future, the expected sensitivities for BR $(\mu\rightarrow e\gamma)$ would be of order $10^{-13}$ \cite{MEG}. For BR $(\tau\rightarrow e\gamma)$ and BR $(\tau\rightarrow\mu\gamma)$, it would be $10^{-9}$ \cite{Bona}. For CR $(\mu-e, Ti)$ in nuclei, it would be as low as $10^{-16}\sim10^{-17}$ \cite{AIP}. Thus, the $\mu-e$ conversion experiments would represent the most promising tool to probe new physics. The study of LFV decays via vector mesons is also waiting for the new data from the experiment.

\section*{Acknowledgements}

One of the authors (KSS) thanks to Prof. J. Rosiek and A. Dedes for providing the fortran codes.
The work has been supported by the National Natural Science Foundation of China (NNSFC)
with Grants No. 10975027 and 11047002.

\appendix

\section{The simplified amplitude in seesaw extended MSSM}\label{app}

In this appendix we present the simplified amplitudes in Fig.\ref{fig1}:
\begin{eqnarray*}
&&{\cal A}_a=\frac{i e^2 \pi ^2 f_{\phi }}{3 m_{\phi } N_c}
(p_3+p_4)\cdot \varepsilon (p)\sum_{i,j=1}^{6}\sum_{k=1}^{4}
\nonumber\\
&&\times\bar{u}_\mu(p_4)\Big\{C_1 m_e(A_{3}^{kj}A_{1}^{ik*}P_L+A_{4}^{kj}A_{2}^{ik*}P_R)
\nonumber\\
&&+C_2\Big[(m_eA_{3}^{kj}A_{1}^{ik*}-m_{\mu }A_{2}^{ik*}A_{4}^{kj}) P_L
\nonumber\\
&&+(m_eA_{2}^{ik*}A_{4}^{kj}-m_{\mu}A_{3}^{kj}A_{1}^{ik*})P_R\Big]
\nonumber\\
&&+C_0\Big[(m_eA_{2}^{ik*}
-m_{\chi_k^0}A_{1}^{ik*})A_{4}^{kj}P_R
\nonumber\\
&&+A_{3}^{kj}(m_eA_{1}^{ik*}-m_{\chi_k^0}A_{2}^{ik*}) P_L\Big]\Big\}\upsilon_e(p_3)
\end{eqnarray*}
where
\begin{eqnarray*}
&&A_{1}^{ik}=\frac{e (s_{\rm w} Z_N^{1k}+c_{\rm w} Z_N^{2 k}) Z_l^{2 i}}{\sqrt{2}
c_{\rm w} s_{\rm w}}+Y_l^2 Z_N^{3 k} Z_l^{5 i}
\nonumber\\
&&A_{2}^{ik}=Y_l^2 Z_l^{2 i} (Z_N^{3 k})^*-\frac{\sqrt{2} e Z_l^{5 i}(Z_N^{1k})^*}{c_{\rm w}}
\nonumber\\
&&A_{3}^{kj}=\frac{e (s_{\rm w} Z_N^{1k}+c_{\rm w} Z_N^{2 k}) Z_l^{1j}}{\sqrt{2}
c_{\rm w} s_{\rm w}}+Y_l^{1}Z_N^{3 k} Z_l^{4 j}
\nonumber\\
&&A_{4}^{kj}=Y_l^{1} Z_N^{3 k} (Z_l^{1j})^*-\frac{\sqrt{2} e Z_N^{1k} (Z_l^{4j})^*}{c_{\rm w}}
\end{eqnarray*}
and
\begin{eqnarray*}
&&C_{0}=C_{0}[m^2,m_{e}^2,m_{\mu}^2,m_{L_{i}}^2,m_{L_{j}}^2,m_{\chi_{k}^0}^2]\\
&&C_{i}=C_{i}[m_{\mu}^2,m_e^2,m_{\phi}^2,m_{L_{i}}^2,m_{\chi_{k}^0}^2,m_{L_{j}}^2]
\end{eqnarray*}
with $i=1,\;2$.
\begin{eqnarray*}
&&{\cal A}_b=\frac{i e^2\pi ^2f_{\phi }m_{\phi} (4 s_{\rm w}^{2}-3)}
{24 N_c s_{\rm w}^{2}c_{\rm w}^{2} (m_{\phi }^2-m_z^2) }(p_3+p_4)\cdot \varepsilon(p)
\sum_{i,j=1}^{6}\sum_{k=1}^{4}
\nonumber\\
&&\times A_{5}^{ij}\bar{u}_\mu(p_4)
\Big\{C_1 m_e (A_{3}^{kj}A_{1}^{ik*}P_L +A_{4}^{kj}A_{2}^{ik*}P_R )
\nonumber\\
&&+C_2\Big[(m_eA_{3}^{kj}A_{1}^{ik*}-m_{\mu}A_{4}^{kj}A_{2}^{ik*})P_L
\nonumber\\
&&+(m_eA_{4}^{kj}A_{2}^{ik*}-m_{\mu}A_{3}^{kj}A_{1}^{ik*})P_R\Big]
\nonumber\\
&&+C_0\Big[(m_eA_{4}^{kj}A_{2}^{ik*}
-m_{\chi_k^0}A_{4}^{kj}A_{1}^{ik*}) P_R
\nonumber\\
&&+(m_eA_{3}^{kj}A_{1}^{ik*}-m_{\chi_k^0}A_{3}^{kj}A_{2}^{ik*})P_L\Big]\Big\}\upsilon_e(p_3)\;,
\end{eqnarray*}
where
\begin{eqnarray*}
&&A_{5}^{ij}=Z_l^{2i}Z_l^{1j\ast}-2 \delta ^{ij} s_{\rm w}^{2}
\end{eqnarray*}
and
\begin{eqnarray*}
&&C_{0}=C_{0}[m^2,m_{e}^2,m_{\mu}^2,m_{L_{i}}^2,m_{L_{j}}^2,m_{\chi_{k}^0}^2]\nonumber\\
&&C_{i}=C_{i}[m_{\mu}^2,m_e^2,m_{\phi}^2,m_{L_{i}}^2,m_{\chi_{k}^0}^2,m_{L_{j}}^2]
\end{eqnarray*}
with $i=1,\;2$.
\begin{eqnarray*}
&&{\cal A}_c=\frac{ie^4 \pi ^2 f_{\phi}m_{\phi}(4 s_{\rm w}^{2}-3)}{48 \sqrt{2}c_{\rm w}^{4}(m_{\phi
}^2-m_z^2)N_c s_{\rm w}^{4}}\sum_{i,j=1}^{4}\sum_{k=1}^{6}
\nonumber\\
&&\times\bar{u}_\mu(p_4)\Big\{2/\!\!\!\varepsilon (p)\Big[\Big(m_em_{\mu}A_{6}^{ji}A_{7}^{jk}(C_0
+C_1+C_2)
\nonumber\\
&&+2s_{\rm w}A_{8}^{jk}\{m_em_{\chi^{0}_{j}}A_{6}^{ji}(C_0+C_1)-m_em_{\chi^{0}_{i}}A_{6}^{ij}C_1\}
\Big)A_{9}^{ik*}
\nonumber\\
&&+2s_{\rm w}A_{10}^{ik*}\Big(A_{7}^{jk}\{A_{6}^{ji}m_{\chi^{0}_{i}}C_0+(A_{6}^{ji}m_{\chi^{0}_{i}}
-A_{6}^{ij}m_{\chi^{0}_{j}}) C_2\}m_{\mu }
\nonumber\\
&&+A_{8}^{jk}\{(4 A_{6}^{ij}-2A_{6}^{ji}m_{\chi^{0}_{i}} m_{\chi^{0}_{j}}) C_0
-A_{6}^{ij}s_{\rm w}[1+2 C_{12}m_{\phi }^2
\nonumber\\
&&-2 (C_1+C_{12}+C_{11})m_{\mu }^2-2 (C_2+C_{12}+C_{22}) m_e^2]\}\Big)\Big]P_R
\nonumber\\
&&+2(p_4\cdot\varepsilon(p))\Big[\Big(A_{6}^{ji}A_{7}^{jk}\{(C_1+C_{11}) m_e P_L+C_{12} m_{\mu } P_R\}
\nonumber\\
&&+2A_{6}^{ij}A_{8}^{jk}s_{\rm w} m_{\chi^{0}_{i}} C_1 P_R\Big)A_{9}^{ik*}
-2 s_{\rm w}\Big(A_{6}^{ji}A_{7}^{jk}m_{\chi^{0}_{i}} C_1 P_L
\nonumber\\
&&+2A_{6}^{ij}A_{8}^{jk}s_{\rm w}\{C_{12} m_{\mu } P_L+(C_1+C_{11}) m_e P_R\}\Big)A_{10}^{ik*}\Big]
\nonumber\\
&&+2(p_3\cdot\varepsilon (p))\Big[2A_{6}^{ij}s_{\rm w}\Big(A_{7}^{jk}m_{\chi^{0}_{j}} C_2 P_L
+2A_{8}^{jk}s_{\rm w}\{(C_2
\nonumber\\
&&+C_{22}) m_{\mu } P_L+C_{12} m_e P_R\}\Big)A_{10}^{ik*}-A_{6}^{ji}\Big(A_{7}^{jk}C_{12} m_e P_L
\nonumber\\
&&+\{A_{7}^{jk}(C_2+C_{22}) m_{\mu }+2A_{8}^{jk}s_{\rm w}m_{\chi^{0}_{j}}C_2\}P_R\Big)A_{9}^{ik*}\Big]
\nonumber\\
&&-/\!\!\!\varepsilon (p)\Big[\Big(A_{6}^{ji}A_{7}^{jk}[(C_2+C_{12}+C_{22}) m_e^2-C_{12}m_{\phi}^2
+(C_1
\nonumber\\
&&+C_{11}+C_{12}) m_{\mu }^2]\Big)+2\Big(\{2A_{6}^{ji}-A_{6}^{ij}\}m_{\chi^{0}_{i}}m_{\chi^{0}_{j}}C_0
\nonumber\\
&&+4A_{8}^{jk}\{A_{6}^{ij}m_{\chi^{0}_{i}} (C_0+C_2)-A_{6}^{ji}m_{\chi^{0}_{j}} C_2\} m_{\mu } s_{\rm w}\Big)A_{9}^{ik*}
\nonumber\\
&&+4 m_e s_{\rm w}\Big(A_{6}^{ij}A_{7}^{jk}m_{\chi^{0}_{j}} C_0+A_{7}^{jk}(A_{6}^{ij}m_{\chi^{0}_{j}}
-A_{6}^{ji}m_{\chi^{0}_{i}}) C_1
\nonumber\\
&&+2A_{6}^{ij}A_{8}^{jk}(C_0+C_1+C_2) m_{\mu } s_{\rm w}\Big)A_{10}^{ik*}-1\Big]P_L
\Big\}\upsilon_e(p_3)
\end{eqnarray*}
where
\begin{eqnarray*}
&&A_{6}^{ij}=Z_{N}^{4i\ast}Z_{N}^{4j}-Z_{N}^{3i\ast}Z_{N}^{3j}\\
&&A_{7}^{jk}=Z_{L}^{1k}(Z_{N}^{1j}s_{\rm w}+Z_{N}^{2j}c_{\rm w})+Y_{l}^{1}Z_{L}^{4k}Z_{N}^{3j}\\
&&A_{8}^{jk}=Z_{L}^{4k}Z_{N}^{1j\ast}+Y_{l}^{1}Z_{L}^{1k}Z_{N}^{3j\ast}\\
&&A_{9}^{ik}=Z_{L}^{2k}(Z_{N}^{1i}s_{\rm w}+Z_{N}^{2i\ast}c_{\rm w})+Y_{l}^{1}Z_{L}^{5k}Z_{N}^{3i}\\
&&A_{10}^{ik}=Z_{L}^{5k}Z_{N}^{1i\ast}+Y_{l}^{2}Z_{L}^{2k}Z_{N}^{3i\ast}.
\end{eqnarray*}
and
\begin{eqnarray*}
&&C_{0}=C_{0}[m_{\phi}^2,m_{e}^2,m_{\mu}^2,m_{\chi{_{i}}}^2,m_{\chi{_{j}^0}}^2,m_{L_{k}}^2]\\
&&C_{{i,ij}}=C_{{i,ij}}[m_{\mu}^2,m_{\phi}^2,m_e^2,m_{L_{k}}^2,m_{\chi_{i}^0}^2,m_{\chi_{j}}^2]
\end{eqnarray*}
with $i=1,\;2$.
\begin{eqnarray*}
&&{\cal A}_d=\frac{i e^2 \pi ^2 f_{\phi }}{6 m_{\phi } N_c s_{\rm w}^{2}}
\sum_{i,j=1}^{2}\sum_{k=1}^{6}
\nonumber\\
&&\times\bar{u}_\mu(p_4)\Big\{/\!\!\!\varepsilon (p)\Big[
A_{11}^{ik*}\Big(\{2(C_2+C_{22})m_e^2
+2(C_1+C_{11})m_{\mu}^2
\nonumber\\
&&+2 C_{12}(m_e^2+m_{\mu }^2-m_{\phi}^2)
+2 C_0 (m_{\chi_i^\pm} m_{\chi_j^\pm}+2)-1\}e^2A_{13}^{jk}
\nonumber\\
&&+2es_{\rm w}A_{14}^{jk}m_{\mu }\{(C_0+C_2) m_{\chi_i^\pm}+C_2 m_{\chi _j^\pm}\}\Big)
+2 m_eA_{12}^{ik*}\nonumber\\
&&\times \Big(es_{\rm w}A_{13}^{jk}\{C_1 m_{\chi _i^\pm}
+(C_0+C_1) m_{\chi_j^\pm}\}-s_{\rm w}^2A_{14}^{jk}(C_0+C_1
\nonumber\\
&&+C_2) m_{\mu }\Big)\Big]P_L
+/\!\!\!\varepsilon (p)\Big[A_{12}^{ik*}\Big(2es_{\rm w}A_{13}^{jk}m_{\mu }\{(C_0+C_2) m_{\chi _i^\pm}
\nonumber\\
&&+C_2 m_{\chi _j^\pm}\}
+s_{\rm w}^2A_{14}^{jk}\{2(C_2+C_{22}) m_e^2+2(C_1+C_{11}) m_{\mu }^2
\nonumber\\
&&+2 C_{12}(m_e^2+m_{\mu}^2-m_{\phi }^2)+2 C_0 (m_{\chi_i^\pm} m_{\chi_j^\pm}+2)-1\}\Big)
\nonumber\\
&&-2m_eA_{11}^{ik*}\Big(e^2A_{13}^{jk}(C_0+C_1+C_2) m_{\mu }
-es_{\rm w}A_{14}^{jk}\{C_1 m_{\chi_i^\pm}
\nonumber\\
&&+(C_0+C_1) m_{\chi _j^\pm}\}\Big)\Big]P_R
-4(p_4\cdot\varepsilon (p))\Big[A_{11}^{ik*}\Big(e^2A_{13}^{jk}\{(C_1
\nonumber\\
&&+C_{11})m_e P_L+C_{12} m_{\mu}P_R\}
+es_{\rm w}A_4 C_1 m_{\chi_i^\pm}P_R\Big)+s_{\rm w}A_{12}^{ik*}
\nonumber\\
&&\times\Big(eA_{13}^{jk}C_1 m_{\chi _i^\pm} P_L
+s_{\rm w}A_{14}^{jk}\{C_{12} m_{\mu }P_L
+C_{11} m_e P_R\}\Big)\Big]
\nonumber\\
&&-(p_3\cdot \varepsilon (p))\Big[A_{11}^{ik*}\Big(e^2A_{13}^{jk}\{C_{12} m_e P_L+(C_2+C_{22})m_{\mu}P_R\}
\nonumber\\
&&-es_{\rm w}A_{14}^{jk}C_2 m_{\chi _j^\pm}P_R\Big)+A_{12}^{ik*}\Big(s_{\rm w}^2A_{14}^{jk}\{(C_2+C_{22})m_{\mu}P_L
\nonumber\\
&&+C_{12}m_e P_R\}-es_{\rm w}A_{13}^{jk}C_2
m_{\chi_j^\pm}P_L\Big)\Big]\Big\}\upsilon_e(p_3)
\end{eqnarray*}
where
\begin{eqnarray*}
&&A_{11}^{ik}=Z_{+}^{1i}(Z_{\nu}^{2k}-I Z_{\nu}^{5k})\\
&&A_{12}^{ik}=Y_{l}^{2}Z_{-}^{2i\ast}(Z_{\nu}^{2k}-I Z_{\nu}^{5k})\\
&&A_{13}^{jk}=Z_{+}^{1j}(Z_{\nu}^{1k}-I Z_{\nu}^{4k})\\
&&A_{14}^{jk}=Y_{l}^{1}Z_{-}^{2j\ast}(Z_{\nu}^{1k}-I Z_{\nu}^{4k})
\end{eqnarray*}
and
\begin{eqnarray*}
&&C_{0}=C_{0}[m_{\phi}^2,m_{e}^2,m_{\mu}^2,m_{\chi{_{i}^\pm}}^2,m_{\chi_{j}^\pm}^2,m_{\nu{_{k}}}^2]\\
&&C_{{i,ij}}=C_{{i,ij}}[m_{\mu}^2,m_{\phi}^2,m_e^2,m_{\nu{_{k}}}^2,m_{\chi_{i}^\pm}^2,m_{\chi_{j}^\pm}^2]
\end{eqnarray*}
with $i=1,2$.
\begin{eqnarray*}
&&{\cal A}_e=\frac{i\pi^2e^2f_{\phi }m_{\phi}(4s_{\rm w}^{2}-3)}{48c_{\rm w}^{2}(m_{\phi }^2-m_z^2)N_c s_{\rm w}^{4}}
\sum_{i,j=1}^{2}\sum_{k=1}^{6}
\nonumber\\
&&\times\bar{u}_\mu(p_4)\Big\{/\!\!\!\varepsilon(p)\Big[s_{\rm w}A_{12}^{ik*}\Big(2eA_{13}^{jk}(A_{16}^{ij} m_{\chi_i^\pm}
+A_{15}^{ij} m_{\chi_j^\pm})C_2 m_{\mu}
\nonumber\\
&&+s_{\rm w}A_{14}^{jk} A_{15}^{ij}\{2 C_2 m_e^2+2 C_{22} m_e^2+2(C_1+C_{11}) m_{\mu}^2-1+2 C_{12}
\nonumber\\
&&\times (m_e^2+m_{\mu }^2-m_{\phi }^2)\}\Big)
-2em_eA_{11}^{ik*}\Big(eA_{13}^{jk}A_{16}^{ij}
(C_1+C_2) m_{\mu }
\nonumber\\
&&-s_{\rm w}A_{14}^{jk}(A_{15}^{ij} m_{\chi _i^\pm}+A_{16}^{ij} m_{\chi _j^\pm})C_1\Big)
+2\Big(s_{\rm w}A_{12}^{ik*}\{A_{13}^{jk} A_{16}^{ij} e m_{\mu } m_{\chi_i^\pm}
\nonumber\\
&&+s_{\rm w}A_{14}^{jk} (2 A_{15}^{ij}+A_{16}^{ij} m_{\chi_i^\pm} m_{\chi _j^\pm})\}
-e m_es_{\rm w}A_{11}^{ik*}A_{16}^{ij}(A_{13}^{jk} e m_{\mu}
\nonumber\\
&&-A_{14}^{jk} m_{\chi _j^\pm})\Big)C_0\Big]P_R
-4(p_4\cdot\varepsilon (p))\Big[eA_{11}^{ik*}\Big(eA_{13}^{jk} A_{16}^{ij}\{(C_1
\nonumber\\
&&+C_{11})
m_e P_L+C_{12} m_{\mu }P_R\}+s_{\rm w}A_{14}^{jk} A_{15}^{ij} C_1 m_{\chi_i^\pm} P_R \Big)
+s_{\rm w}A_{12}^{ik*}
\nonumber\\
&&\times\Big(eA_{13}^{jk}A_{16}^{ij}C_1 m_{\chi_i^\pm} P_L+s_{\rm w}A_{14}^{jk}
A_{15}^{ij}\{(C_1+C_{11}) m_e P_R
\nonumber\\
&&+C_{12} m_{\mu }P_L\}\Big)\Big]
+/\!\!\!\varepsilon (p)\Big[eA_{11}^{ik*}\Big(eA_{13}^{jk} A_{16}^{ij}\{2 (C_2+ C_{22}) m_e^2
\nonumber\\
&&+2 (C_1+C_{11})m_{\mu }^2+2C_{12} (m_e^2+m_{\mu }^2-m_{\phi }^2)-1\}+2 s_{\rm w}A_{14}^{jk}
\nonumber\\
&&\times(A_{15}^{ij}m_{\chi_i^\pm}+A_{16}^{ij} m_{\chi _j^\pm})C_2 m_{\mu }
+2 C_0\{2eA_{13}^{jk} A_{16}^{ij}+A_{15}^{ij} m_{\chi_i^\pm}
\nonumber\\
&&\times (eA_{13}^{jk}m_{\chi _j^\pm}+s_{\rm w}A_{14}^{jk} m_{\mu })\}\Big)
+2 m_e s_{\rm w}A_{12}^{ik*}\Big(eA_{13}^{jk}\{A_{16}^{ij} C_1 m_{\chi_i^\pm}
\nonumber\\
&&+A_{15}^{ij}(C_0+C_1) m_{\chi _j^\pm}\}
-s_{\rm w}A_{14}^{jk} A_{15}^{ij} (C_0+C_1+C_2) m_{\mu }\Big)\Big]P_L
\nonumber\\
&&+4(p_3\cdot \varepsilon (p))\Big[eA_{11}^{ik*}A_{16}^{ij}\Big(eA_{13}^{jk}C_{12} m_e P_L
+\{eA_{13}^{jk}(C_2+C_{22}) m_{\mu }
\nonumber\\
&&- s_{\rm w}A_{14}^{jk} C_2 m_{\chi _j^\pm}\}P_R\Big)
+s_{\rm w}A_{12}^{ik*}A_{15}^{ij}\Big(s_{\rm w}A_{14}^{jk}\{(C_2+C_{22}) m_{\mu } P_L
\nonumber\\
&&+C_{12} m_e P_R\}-eA_{13}^{jk} C_2 m_{\chi _j^\pm} P_L\Big)\Big]\Big\}\upsilon_e(p_3)
\end{eqnarray*}
where
\begin{eqnarray*}
&&A_{15}^{ij}=Z_{+}^{1i\ast}Z_{+}^{1j}+\delta^{ij}(c_{\rm w}^{2}-s_{\rm w}^{2})\\
&&A_{16}^{ij}=Z_{-}^{1i}Z_{-}^{1j\ast}+\delta^{ij}(c_{\rm w}^{2}-s_{\rm w}^{2})
\end{eqnarray*}
and
\begin{eqnarray*}
&&C_{0}=C_{0}[m_{\phi}^2,m_{e}^2,m_{\mu}^2,m_{\chi{_{i}^\pm}}^2,m_{\chi{_{j}^\pm}}^2,m_{\nu{_{k}}}^2]\\
&&C_{{i,ij}}=C_{{i,ij}}[m_{\mu}^2,m_{\phi}^2,m_e^2,m_{\nu{_{k}}}^2,m_{\chi_{i}^\pm}^2,m_{\chi_{j}^\pm}^2]
\end{eqnarray*}
with i=1,2.
\begin{eqnarray*}
&&{\cal A}_f=\frac{i \pi ^2e^2 f_{\phi } m_{\phi } (3-4 s_{\rm w}^{2})}{24 N_c
s_{\rm w}^{4}c_{\rm w}^{2} (m_{\phi }^2-m_z^2) }(p_3+p_4)\cdot\varepsilon (p)
\sum_{i,j=1}^{6}\sum_{k=1}^{2}
\nonumber\\
&&\times\bar{u}_\mu(p_4)
\Big\{C_1m_{\mu }\Big(e^2A_{11}^{ki*}A_{13}^{kj} P_R
+s_{\rm w}^{2}A_{12}^{ki} A_{14}^{kj} P_L\Big)
\nonumber\\
&&+C_2\Big[e^2A_{11}^{ki*}A_{13}^{kj} (m_{\mu } P_R-m_e P_L)
+s_{\rm w}^{2}A_{12}^{ki} A_{14}^{kj}(m_{\mu } P_L
\nonumber\\
&&-m_e P_R)\Big]
+C_0\Big[eA_{11}^{ki*}(A_{13}^{kj}e m_{\mu }
-s_{\rm w}A_{14}^{kj} m_{\chi_{k}^\pm})P_R
\nonumber\\
&&+s_{\rm w} A_{12}^{ki} (s_{\rm w}m_{\mu }A_{14}^{kj}
-e m_{\chi_{k}^\pm}A_{13}^{kj})P_L\Big]\Big\}\upsilon_e(p_3)
\end{eqnarray*}
where
\begin{eqnarray*}
&&C_{0}=C_{0}[m_{\phi}^2,m_{e}^2,m_{\mu}^2,m_{\nu{_{i}}}^2,m_{\nu{_{j}}}^2,m_{\chi_{k}^\pm}^2]\\
&&C_{i}=C_{i}[m_{e}^2,m_{\mu}^2,m_{\phi}^2,m_{\nu{_{i}}}^2,m_{\chi_{k}^\pm}^2,m_{\nu_{j}}^2],i=1,2
\end{eqnarray*}
the simplified amplitudes in Fig.\ref{fig2}:
\begin{eqnarray*}
&&{\cal A}_g=\frac{i \pi ^2}{48 N_{c} s_{\rm w}^2} \sum_{i,m=1}^{2}\sum_{j=1}^{6}
\sum_{l=1}^{6}
\nonumber\\
&&\times\bar{u}_\mu(p_4)\Big\{2f_{\phi} m_{\phi} \Big[A_{13}^{mj}
eD_{0}A_{17}^{ml\ast} A_{17}^{il}(em_{e} m_{\mu}A_{11}^{ij\ast}
\nonumber\\
&&+s_{\rm w}m_{e} m_{\chi_{i}^\pm}A_{12}^{ij\ast})-A_{14}^{mj}s_{\rm w}D_{0}A_{17}^{ml\ast}
A_{17}^{il}(e m_{\chi_{m}^\pm} m_{\mu}A_{11}^{ij\ast}
\nonumber\\
&&+s_{\rm w}m_{\chi_{i}^\pm}m_{\chi_{m}^\pm}A_{12}^{ij\ast})
-A_{14}^{mj} s_{\rm w}^2A_{12}^{ij\ast}A_{18}^{ml\ast} A_{18}^{il} (2 C_{0}
\nonumber\\
&&+2 D_{1} m_{\nu_{j}}^2)\Big]/\!\!\!\varepsilon (p)P_R
+2f_{\phi}m_{\phi}\Big[-A_{13}^{mj}e^2A_{11}^{ij\ast}A_{17}^{ml\ast} A_{17}^{il}
\nonumber\\
&&\times(2 C_{0}+2 D_{1} m_{\nu_{j}}^2)
+A_{14}^{mj}D_{0} A_{18}^{ml\ast} A_{18}^{il}\{s_{\rm w}(em_{e}
m_{\chi_{i}^\pm}
\nonumber\\
&&\times A_{11}^{ij\ast}
+m_{e} m_{\mu}s_{\rm w}A_{12}^{ij\ast})
-e(m_{\chi_{m}^\pm}m_{\mu}s_{\rm w}A_{12}^{ij\ast}
+em_{\chi_{i}^\pm}
\nonumber\\
&&\times m_{\chi_{m}^\pm}A_{11}^{ij\ast})\}\Big]/\!\!\!\varepsilon (p)P_L
+D_{0} f^{T}_{\phi}A_{17}^{ml\ast}A_{18}^{il}\Big[eA_{13}^{mj}(em_{e}
\nonumber\\
&&\times m_{\chi_{i}^\pm}A_{11}^{ij\ast}
+m_{e} m_{\mu}s_{\rm w}A_{12}^{ij\ast})
-A_{14}^{mj} s_{\rm w} (e m_{\chi_{i}^\pm} m_{\chi_{m}^\pm}A_{11}^{ij\ast}
\nonumber\\
&&+m_{\chi_{m}^\pm}
m_{\mu} s_{\rm w}A_{12}^{ij\ast})\Big]
(/\!\!\!\varepsilon (p)/\!\!\!p-/\!\!\!p/\!\!\!\varepsilon (p))P_R
+D_{0} f^{T}_{\phi}A_{18}^{ml\ast} A_{17}^{il}
\nonumber\\
&&\times\Big[s_{\rm w}A_{14}^{mj}(em_{e}
m_{\mu}A_{11}^{ij\ast}+m_{e}m_{\chi_{i}^\pm}s_{\rm w}A_{12}^{ij\ast})
-eA_{13}^{mj}
\nonumber\\
&&\times (e m_{\chi_{m}^\pm} m_{\mu}A_{11}^{ij\ast}+m_{\chi_{i}^\pm}m_{\chi_{m}^\pm}s_{\rm w}A_{12}^{ij\ast})\Big]
(/\!\!\!\varepsilon (p)/\!\!\!p
-/\!\!\!p/\!\!\!\varepsilon (p))
\nonumber\\
&&\cdot P_L\Big\}\upsilon_e(p_3)
\end{eqnarray*}
where
\begin{eqnarray*}
&&A_{17}^{(i,m)l}=\Big[-\frac{e}{s_{\rm w}}Z_{U}^{Jl\ast}Z_{+}^{1(i,m)}+Y_{u}^{J}Z_{U}^{(J+3)l}Z_{+}^{2(i,m)}\Big]K^{J2}\\
&&A_{18}^{(i,m)l}=\Big[-Y_{d}^{2}Z_{U}^{Jl\ast}Z_{-}^{2(i,m)\ast}\Big]K^{J2}
\end{eqnarray*}
and
\begin{eqnarray*}
&&C_{0}=C_{0}[m_{\phi}^2,m_{s}^2,m_{s}^2,m_{\chi{_{i}^\pm}}^2,m_{\chi_{j}^\pm}^2,m_{U_{l}}^2]\\
&&D_{0}=D_{0}[m_{e}^2,m_{s}^2,m_{s}^2,m_{\mu}^2,m_{D}^2,m_{\phi}^2,m_{\nu_{j}}^2,m_{\chi{_{m}^\pm}}^2,m_{U_{l}}^2,m_{\chi_{i}^\pm}^2]\\
&&D_{1}=D_{0}[m_{D}^2,m_{s}^2,m_{\phi}^2,m_{\mu}^2,m_{e}^2,m_{s}^2,m_{\nu_{j}}^2,m_{U_{l}}^2,m_{\chi{_{m}^\pm}}^2,m_{\chi_{i}^\pm}^2]
\end{eqnarray*}
where $m_{D}^2=m_{\mu}^2+m_{s}^2-0.5m_{\phi}^2$
\begin{eqnarray*}
&&{\cal A}_h=\frac{i \pi ^2 }{48 N_c}\sum_{i,m=1}^{4}\sum_{j,l=1}^{6}
\nonumber\\
&&\times\bar{u}_\mu(p_4)\Big\{2 f_{\phi} m_{\phi} \Big[A_{2}^{ji\ast} (A_{19}^{ml\ast} D_0 A_{19}^{il}
   m_{\chi{_{i}^0}} (A_{3}^{mj} m_{e}
\nonumber\\
&&-A_{4}^{mj} m_{\chi{_{m}^0}})-A_{20}^{ml\ast} A_{20}^{il} (2 C_0
   A_{4}^{mj}+2 D_1 A_{4}^{mj} m_{L_{j}}^2))
\nonumber\\
&&+A_{1}^{ji\ast}D_0A_{19}^{ml\ast} A_{19}^{il} (A_{3}^{mj} m_{e} m_{\mu}-
   A_{4}^{mj} m_{\chi{_{m}^0}} m_{\mu})\Big] /\!\!\!\varepsilon (p)P_R
\nonumber\\
&&+2 f_{\phi} m_{\phi}
   \Big[A_{2}^{ji\ast}D_0A_{20}^{ml\ast} A_{20}^{il} (A_{4}^{mj} m_{e} m_{\mu}-A_{3}^{mj} m_{\chi{_{m}^0}}
   m_{\mu})
\nonumber\\
&&+A_{1}^{ji\ast} (A_{20}^{ml\ast} D_0 A_{20}^{il} m_{\chi{_{i}^0}} (A_{4}^{mj}m_{e}-A_{3}^{mj} m_{\chi{_{m}^0}})
-2A_{19}^{ml\ast}
\nonumber\\
&&\times A_{19}^{il} (C_0 A_{3}^{mj}+D_1A_{3}^{mj} m_{L_{j}}^2))\Big] /\!\!\!\varepsilon (p)P_L
+D_0 f_{\phi}^{T}A_{19}^{ml\ast}
\nonumber\\
&&\times A_{20}^{il}(A_{1}^{ji\ast}m_{\chi{_{i}^0}}+A_{2}^{ji\ast} m_{\mu}) (A_{3}^{mj} m_{e}-A_{4}^{mj} m_{\chi{_{m}^0}})
(/\!\!\!\varepsilon (p)/\!\!\!p
\nonumber\\
&&-/\!\!\!p/\!\!\!\varepsilon (p))P_R
+D_0 f_{\phi}^{T}A_{20}^{ml\ast} A_{19}^{il}(A_{1}^{ji\ast} m_{\mu}+A_{2}^{ji\ast}m_{\chi{_{i}^0}})
\nonumber\\
&&\times (A_{4}^{mj} m_{e}-A_{3}^{mj} m_{\chi{_{m}^0}})
(/\!\!\!\varepsilon (p)/\!\!\!p-/\!\!\!p/\!\!\!\varepsilon (p))P_L\Big\}\upsilon_e(p_3)
\end{eqnarray*}
where
\begin{eqnarray*}
&&A_{19}^{(i,m)l}=-\frac{e}{s_{\rm w}}Z_{D}^{2l}(\frac{s_{\rm w}}{3}Z_{N}^{1(i,m)}-Z_{N}^{2(i,m)}c_{\rm w})+Y_{d}^{2}Z_{D}^{5l}Z_{N}^{3(i,m)}\\
&&A_{20}^{(i,m)l}=-\frac{\sqrt{2}e}{3c_{\rm w}}Z_{D}^{5l}Z_{N}^{1(i,m)\ast}+Y_{D}^{2}Z_{D}^{2l}Z_{N}^{3(i,m)\ast}
\end{eqnarray*}
and
\begin{eqnarray*}
&&C_{0}=C_{0}[m_{\phi}^2,m_{s}^2,m_{s}^2,m_{\chi{_{i}^0}}^2,m_{\chi_{m}^0}^2,m_{D_{l}}^2]\\
&&D_{0}=D_{0}[m_{e}^2,m_{s}^2,m_{s}^2,m_{\mu}^2,m_{D}^2,m_{\phi}^2,m_{L_{j}}^2,m_{\chi{_{m}^0}}^2,m_{D_{l}}^2,m_{\chi_{i}^0}^2]\\
&&D_{1}=D_{0}[m_{D}^2,m_{s}^2,m_{\phi}^2,
m_{\mu}^2,m_{e}^2,m_{s}^2,m_{L_{j}}^2,m_{D_{l}}^2,m_{\chi{_{m}^0}}^2,m_{\chi_{i}^0}^2]
\end{eqnarray*}
where $m_{D}^2=m_{\mu}^2+m_{s}^2-0.5m_{\phi}^2$
\begin{eqnarray*}
&&{\cal A}_i=\frac{i \pi ^2 }{48 N_c}\sum_{i,m=1}^{4}\sum_{j=1,l}^{6}
\nonumber\\
&&\times\bar{u}_e(p_3)\Big\{2 f_{\phi} m_{\phi} \Big[A_{4}^{ij\ast} (A_{19}^{ml\ast} D_0 A_{19}^{il}
   m_{\chi{_{i}^0}} (A_{1}^{jm} m_{e}
\nonumber\\
&&-A_{2}^{jm} m_{\chi{_{m}^0}})-A_{20}^{ml\ast} A_{20}^{il} (2 C_0
   A_{2}^{jm}+2 D_1 A_{2}^{jm} m_{L_{j}}^2))
\nonumber\\
&&+A_{3}^{ij\ast}D_0A_{19}^{ml\ast} A_{19}^{il} (A_{1}^{jm} m_{e} m_{\mu}-
   A_{2}^{jm} m_{\chi{_{m}^0}} m_{\mu})\Big] /\!\!\!\varepsilon (p)P_R
\nonumber\\
&&+2 f_{\phi} m_{\phi}
   \Big[A_{4}^{ij\ast}D_0A_{20}^{ml\ast} A_{20}^{il} (A_{2}^{jm} m_{e} m_{\mu}-A_{1}^{jm} m_{\chi{_{m}^0}}
   m_{\mu})
\nonumber\\
&&+A_{3}^{ij\ast} (A_{20}^{ml\ast} D_0 A_{20}^{il} m_{\chi{_{i}^0}} (A_{2}^{jm}m_{e}-A_{1}^{jm} m_{\chi{_{m}^0}})
-2A_{19}^{ml\ast}
\nonumber\\
&&\times A_{19}^{il} (C_0 A_{1}^{jm}+D_1A_{1}^{jm} m_{L_{j}}^2))\Big] /\!\!\!\varepsilon (p)P_L
+D_0 f_{\phi}^{T}A_{19}^{ml\ast}
\nonumber\\
&&\times A_{20}^{il}(A_{3}^{ij\ast}m_{\chi{_{i}^0}}+A_{4}^{ij\ast} m_{\mu})
(A_{1}^{jm} m_{e}-A_{2}^{jm} m_{\chi{_{m}^0}})
\nonumber\\
&&(/\!\!\!\varepsilon (p)/\!\!\!p-/\!\!\!p/\!\!\!\varepsilon (p))P_R
+D_0 f_{\phi}^{T}A_{20}^{ml\ast} A_{19}^{il}(A_{3}^{ij\ast} m_{\mu}+A_{4}^{ij\ast}
\nonumber\\
&&\times m_{\chi{_{i}^0}})
(A_{2}^{jm} m_{e}-A_{1}^{jm} m_{\chi{_{m}^0}})
(/\!\!\!\varepsilon (p)/\!\!\!p-/\!\!\!p/\!\!\!\varepsilon (p))P_L\Big\}
\upsilon_\mu(p_4)
\end{eqnarray*}
where
\begin{eqnarray*}
&&C_{0}=C_{0}[m_{\phi}^2,m_{s}^2,m_{s}^2,m_{\chi{_{i}^0}}^2,m_{\chi_{m}^0}^2,m_{D_{l}}^2]\\
&&D_{0}=D_{0}[m_{\mu}^2,m_{s}^2,m_{s}^2,m_{e}^2,m_{D}^2,m_{\phi}^2,m_{L_{j}}^2,m_{\chi{_{m}^0}}^2,m_{D_{l}}^2,m_{\chi_{i}^0}^2]\\
&&D_{1}=D_{0}[m_{D}^2,m_{s}^2,m_{\phi}^2,m_{\mu}^2,m_{e}^2,m_{s}^2,m_{L_{j}}^2,m_{D_{l}}^2,m_{\chi{_{i}^0}}^2,m_{\chi_{m}^0}^2]
\end{eqnarray*}
where $m_{D}^2=m_{e}^2+m_{s}^2-0.5m_{\phi}^2$

\section*{References}

\bibliographystyle{model1-num-names}

\end{document}